\documentclass[fleqn,usenatbib]{mnras}
\usepackage{newtxtext,newtxmath} 
\usepackage{ae,aecompl}
\usepackage{graphicx}	% Including figure files
\usepackage{amsmath}	% Advanced maths commands
\usepackage[dvipsnames]{xcolor}  
\usepackage{amssymb}	% Extra maths symbols
\usepackage{hyperref}
\usepackage[normalem]{ulem}
\usepackage{soul}
\usepackage{macros_vdb} % my personal macros
\begin{document}
%%%%%%%%%%%%%%%%%%%%%%%%%%%%%%%%%%%%%%%%%%%%%%%%%%%%%%%%%%%%%%%%%%%%%%%%%%

\title[DASH library]{DASH: a library of dynamical subhalo evolution}

\author[G. Ogiya et al.]{%
    Go Ogiya$^{1}$\thanks{E-mail: go.ogiya@oca.eu (GO)},
    Frank~C.~van den Bosch$^{2}$,
    Oliver Hahn$^{1}$, 
    Sheridan~B.~Green$^{3}$,
    \newauthor
    Tim~B.~Miller$^{2}$
    and Andreas Burkert$^{4,5}$
\vspace*{8pt}
\\
  $^{1}$Universit\'e C\^ote d'Azur, Observatoire de la C\^ote d'Azur, CNRS, Laboratoire Lagrange\\ \quad Blvd de l'Observatoire, CS 34229, 06304 Nice, France \\
  $^{2}$Department of Astronomy, Yale University, PO. Box 208101, New Haven, CT 06520-8101 \\
  $^{3}$Department of Physics, Yale University, PO. Box 208120, New Haven, CT 06520-8120 \\
  $^{4}$Universit\"ats-Sternwarte M\"unchen, Scheinerstra\ss e 1, D-81679 M\"unchen, Germany \\
  $^{5}$Max-Planck-Institut f\"ur extraterrestrische Physik, Postfach 1312, Gie\ss enbachstra\ss e, D-85741 Garching, Germany 
}

%%%%%%%%%%%%%%%%%%%%%%%%%%%%%%%%%%%%%%%%%%%%%%%%%%%%%%%%%%%%%%%%%%%%%%%%%%

\date{}

\pagerange{\pageref{firstpage}--\pageref{lastpage}}
\pubyear{2013}

\maketitle

\label{firstpage}

%%%%%%%%%%%%%%%%%%%%%%%%%%%%%%%%%%%%%%%%%%%%%%%%%%%%%%%%%%%%%%%%%%%%%%%%%%

\begin{abstract}
  The abundance and demographics of dark matter substructure is important for many areas in astrophysics and cosmological $N$-body simulations have been the primary tool used to investigate them. However, it has recently become clear that the simulations are subject to numerical artefacts, which hampers a proper treatment of the tidal evolution of subhaloes. Unfortunately, no analytical models that accurately describe subhalo evolution exist either. We therefore present a library of idealized, high resolution $N$-body simulations of the tidal evolution of individual subhaloes that can be used to calibrate semi-analytical models and to complement cosmological simulations. The simulations focus on minor mergers, i.e., the mass of the subhalo is much smaller than that of the host halo, such that the impact of dynamical friction is negligible. This setup allows the adoption of a fixed analytical potential for modelling the host halo. The dynamical evolution of subhaloes is followed with $N$-body computations. In the library, four parameters, two of which characterize the subhalo orbit with respect to the host halo, and the two concentrations of the host- and subhalo, are varied over the ranges encountered in cosmological simulations. We show several representative examples from the library that illustrate the evolution of the subhalo mass and velocity dispersion profiles. Additionally, we make publicly available a pre-trained non-parametric model of the subhalo mass evolution based on random forest regression. This model is able to interpolate the simulation data at the 0.1\,dex level and provides efficient access to the data for further use in modelling\footnotemark.
\end{abstract}

%%%%%%%%%%%%%%%%%%%%%%%%%%%%%%%%%%%%%%%%%%%%%%%%%%%%%%%%%%%%%%%%%%%%%%%%%%


%%%%%%%%%%%%%%%%%%%%%%%%%%%%%%%%%%%%%%%%%%%%%%%%%%%%%%%%%%%%%%%%%%%%%%%%%%
\footnotetext{DASH simulation data and machine learning model are available at \url{https://cosmo.oca.eu/dash}.} 

%%%%%%%%%%%%%%%%%%%%%%%%%%%%%%%%%%%%%%%%%%%%%%%%%%%%%%%%%%%%%%%%%%%%%%%%%%%%%%%%%%%%
%%%%%%%%%%%%%%%%%%%%%%%%%%%%%%%%%%%%%%%%%%%%%%%%%%%%%%%%%%%%%%%%%%%%%%%%%%%%%%%%%%%%
\section{Introduction}
\label{sec:intro}

In the cold dark matter (CDM) paradigm of cosmic structure formation, smaller perturbations collapse first to form virialized dark matter haloes, leading to a hierarchical assembly of haloes. When dark matter haloes assemble their mass by accreting smaller haloes, they build up a hierarchy of substructure, with subhaloes hosting sub-subhaloes, hosting sub-sub-subhaloes, etc. As these subhaloes orbit their hosts, they experience mass loss due to the combined effect of dynamical friction, tidal stripping and impulsive (tidal) heating \citep[e.g.,][]{Mo2010}. The resulting abundance and demographics of substructure depends on the microscopic properties of the dark matter particles, most importantly the free-streaming scale and the strength of dark matter self-interaction \citep[see e.g.,][]{Knebe2008, Rocha.etal.13, Bose.etal.16}. This is why many efforts are underway to quantify the amount of dark matter substructure using, among others, gravitational lensing \citep[e.g.][]{Vegetti2012, Shu2015, Hezaveh2016}, gaps in stellar streams \citep[e.g.][]{Carlberg2012, Ngan2014, Erkal2016}, and annihilation or decay signals of dark matter particles \citep[e.g.][]{Strigari2007, Pieri2008, Hayashi2016, Hiroshima2018}. In addition, substructure is also directly related to the abundance and properties of satellite galaxies \citep[e.g.,][]{Vale.Ostriker.06, Newton.etal.18}, and thus to the clustering amplitude of galaxies on small scales \citep[see e.g.,][]{Benson.etal.01, Berlind.etal.03, Kravtsov.etal.04}. Hence, it is important that we are able to make accurate predictions for the abundance of dark matter substructure for a given cosmological model. Given the highly non-linear nature of the processes involved, this is ideally done using $N$-body simulations.

Modern, state-of-the-art cosmological $N$-body simulations predict that roughly 5-10 percent of a halo's mass is bound up in substructure, with more massive host haloes having a larger subhalo mass fraction \citep[e.g.,][]{Gao.etal.04, Giocoli.etal.10}. In addition, subhaloes are found to be spatially anti-biased with respect to the dark matter, in that the radial number density profile, $n_{\rm sub}(r)$, is less centrally concentrated than the halo density profile \citep[e.g.,][]{Diemand.Moore.Stadel.04, Nagai.Kravtsov.05, Springel.etal.08}. It has also been concluded that the subhalo mass function has a universal form \citep[][]{Jiang2016}, albeit with a significant halo-to-halo variance at the massive end \citep[][]{Jiang.vdBosch.17}.

Although these trends are well understood \citep[][]{vandenBosch2005, Zentner2005, Jiang2016}, and the results from numerical simulations seem to be well converged \citep[e.g.,][]{Springel.etal.08, Onions2012}, some issues remain. Foremost among these is the fact that subhaloes in numerical simulations typically experience complete disruption some time after accretion \citep[e.g.,][]{Han2016, vandenBosch2017}. Although it is often argued that this disruption is a physical consequence of either tidal stripping or tidal heating \citep[][]{Hayashi2003, Taylor.Babul.04, Klypin.etal.15}, others have argued that in a collisionless dark matter simulation, subhaloes should rarely ever completely disrupt. In particular, in \citet[hereafter \paperi{}]{PaperI}, we have demonstrated that neither tidal stripping nor tidal heating is expected to be able to completely unbind the central cusps of CDM substructure. The same conclusion was reached by \cite{Penarrubia2010} using idealized, high-resolution numerical simulations. In \citet[hereafter \paperii{}]{PaperII}, we used a large suite of similar, idealized simulations to demonstrate that the disruption of subhaloes in $N$-body simulations is predominantly numerical, and triggered by two independent aspects: discreteness noise and inadequate force softening.

An important finding of \paperii{} is that this artificial, numerical disruption may elude standard convergence tests, in that the method that is typically used to scale the force softening with the mass resolution is inadequate to overcome these problems. Artificial disruption can potentially have far-reaching consequences. After all, unless we can make accurate, and above all reliable, predictions regarding the abundance and structure of dark matter subhaloes, we will forfeit one of the main handles we have on learning about the nature of dark matter. In addition, artificial disruption is also a serious road-block for the small-scale cosmology program, which often relies on numerical simulations to predict the clustering strength of galaxies on small scales. A prime example of this is subhalo abundance matching, which assigns `mock' galaxies to subhaloes identified in numerical simulations in order to predict galaxy-galaxy correlation functions \citep[e.g.,][]{Vale.Ostriker.06, Conroy.etal.06, Guo2010, Hearin.etal.13}. Although one may overcome the implications of artificial disruption by including `orphan' galaxies (i.e., mock galaxies without an associated subhalo in the simulation), this seriously diminishes the information content of small-scale clustering, unless it is well understood how many orphans to add and where.

Unfortunately, it is not clear how to improve $N$-body codes such that the issue of artificial disruption can be avoided. As a consequence, it is difficult to gauge its potential impact on our predictions for the abundance of substructure. However, some insight can be gained from the semi-analytical models constructed by \cite{Jiang2016}. Using accurate halo merger trees, this model uses a simple, orbit-averaged prescription for mass stripping to predict the evolved subhalo mass function.  The overall normalization for the efficiency of mass stripping is calibrated by matching the model predictions to those from a high-resolution cosmological $N$-body simulation. In addition, the model includes a treatment of subhalo disruption, which is also calibrated to accurately reproduce the disruption in the simulation. If we use this model, but turn off the disruption (rather than disrupting the subhaloes, we continue to strip their mass), the resulting subhalo mass function is roughly a factor of two higher than with disruption (Green et al., in prep). Hence, if all disruption is indeed artificial, and if the mass stripping model used by \citet{Jiang2016} is roughly correct, numerical simulations may have been underpredicting the amount of surviving substructure by a factor of two. This would have far-reaching consequences for many areas of astrophysics. For instance, this factor of two is exactly what is needed to solve the `galaxy clustering crisis' in subhalo abundance matching discussed in \cite{Campbell.etal.18}.

In order to make more reliable predictions, we need to develop more sophisticated semi-analytical models for the evolution of dark matter substructure.  Numerous studies in the past have been devoted to this \citep[e.g.,][]{Taylor2001, Penarrubia.Benson.05, Zentner2005, Diemand2007, Kampakoglou.Benson.07, Gan2010, Pullen.etal.14}, but they all have one shortcoming in common: they all use the outcome of cosmological $N$-body simulations in order to calibrate one or more `fudge' parameters in their model. And in doing so, their semi-analytical models inherit the shortcomings of the simulations; put differently, by construction the models are only as accurate as the simulations used for their calibration.

In an attempt to bypass this shortcoming, this paper presents a large database (called DASH, for Dynamical Aspects of SubHaloes), of more than 2,000 idealized, high-resolution simulations of the tidal evolution of individual subhaloes. The simulations cover most of the parameter space (mass ratios, orbital parameters and halo concentrations) relevant for modelling the tidal evolution of subhaloes as they orbit their hosts, and each simulation is evolved with sufficient numerical resolution that discreteness noise and force softening do not adversely affect their outcome. The primary goal of DASH is to enable a more accurate calibration and validation of analytical treatments of tidal stripping and heating, thereby allowing for the construction of new and improved semi-analytical models for dark matter substructure that are not hampered by artificial disruption. In addition, we provide a non-parametric model of subhalo mass evolution, using random forest regression, along with the simulation data. This model describes the simulation results at the $\sim$\,0.1\,dex level and can be readily used in further modelling.

This paper is organized as follows: \S\ref{sec:overview} describes the simulation setup and gives an overview of the DASH library. \S\ref{sec:examples} presents a few examples, highlighting the type of data that is available. In \S\ref{sec:ml}, we demonstrate and validate the performance of a random forest regression model trained on the simulation data to predict bound mass fractions as a function of time since accretion for given orbital parameters and given concentrations of the sub- and host haloes. Finally, \S\ref{sec:summary} summarizes the paper and discusses the future outlook.

%%%%%%%%%%%%%%%%%%%%%%%%%%%%%%%%%%%%%%%%%%%%%%%%%%%%%%%%%%%%%%%%%%%%%%%%%%%%%%%%%%%%
%%%%%%%%%%%%%%%%%%%%%%%%%%%%%%%%%%%%%%%%%%%%%%%%%%%%%%%%%%%%%%%%%%%%%%%
\section{Overview of the simulations}
\label{sec:overview}

This section presents an overview of the DASH library of idealized, collisionless $N$-body simulations of halo minor mergers. After describing the initial conditions (\S\ref{sssec:ICs}), the simulation code (\S\ref{sssec:numtech}), and the analysis and products (\S\ref{sssec:analysis}), we describe the method used to sample the parameter space (\S\ref{ssec:orb_c_params}) and the data format of the DASH library (\S\ref{ssec:data_str}).

%%%%%%%%%%%%%%%%%%%%%%%%%%%%%%%%%%%%%%%%%%%%%%%%%%%%%%%%%%%%%%%%%%%%%%%%%%%%%%%%%%%%
\subsection{Simulation setup}
\label{ssec:setup}

%%%%%%%%%%%%%%%%%%%%%%%%%%%%%%%%%%%%%%%
\subsubsection{Halo profiles and merger set-up}
\label{sssec:ICs}

Each of the simulations run as part of the DASH library follows an individual $N$-body subhalo as it orbits the fixed, external potential of a host halo. Both the host halo and the {\it  initial} (prior to the onset of tidal stripping) subhalo are assumed to be spherical, and to have a Navarro-Frenk-White \citep[NFW;][]{Navarro1997} density profile
\begin{equation}\label{NFWprof}
  \rho(r) = \rho_0 \, \left(\frac{r}{r_\rms}\right)^{-1} \,
  \left(1 + \frac{r}{r_\rms}\right)^{-2}\,,  
\end{equation}
where $r_\rms$ and $\rho_0$ are the characteristic scale radius and density, respectively. We define the virial radius, $\rvir$, as the radius inside of which the average density is $\Delta_{\rm vir} = 200$ times the critical density given by $\rho_{\rm crit} = (3 \, H^2_0/8 \pi G)$, where $H_0$ and $G$ are the Hubble constant and gravitational constant, respectively. The virial mass of the halo is given by
\begin{equation}
M\sub{vir} = \frac{4 \pi}{3} \, \Delta\sub{vir} \, \rho\sub{crit} \, r^3\sub{vir}\,.
\label{eq:virial_mass}
\end{equation}
We emphasize, though, that the DASH simulations also apply to other values of $\Delta_{\rm vir}$, as detailed in \S\ref{ssec:orb_c_params} and Appendix~\ref{App:invariance}. The halo concentration is defined as $c \equiv \rvir/r_\rms$, and the virial velocity is defined as the circular velocity at the virial radius, $V_{\rm vir} = \sqrt{G \Mvir/\rvir}$. The mass ratio between the host halo and the initial subhalo is specified by $\calM \equiv \Mhost/\Msub$, where, as throughout this paper, subscripts `h' and `s' indicate properties of the host- and subhaloes, respectively.

Initial conditions are generated under the assumption that the NFW subhalo is isolated and has an isotropic velocity distribution, such that its phase-space distribution function (DF) depends only on energy, i.e., $f = f(E)$. We use the method of \cite{Widrow.00} to sample particles from the DF using the standard acceptance-rejection technique \citep[][]{Press.etal.92, Kuijken.Dubinski.94}. When computing the DF, we follow \cite{Kazantzidis2006} and assume that the initial NFW subhalo has an exponentially decaying density profile for $r > \rsub$. We assume that the total mass in this exponential extension is $0.05 \Msub$, where $\Msub = M(< \rsub)$. Requiring a smooth transition in the density profile at $r=\rsub$ then determines the scale radius of the exponential decay. Note, though, that when we sample particles from the DF thus computed, we apply a hard truncation at $r = \rsub$ (i.e., no particles are sampled beyond the subhalo's virial radius). Consequently, the system will deviate somewhat from perfect equilibrium near the truncation radius. However, since we embed the subhalo in an external tidal field, with a corresponding tidal radius that typically lies well inside of $\rsub$, there is little virtue to having a subhalo whose outskirts are in perfect equilibrium. In fact, one might argue that it is more realistic to truncate the subhalo at the tidal radius corresponding to its initial position within the host halo. However, in addition to the tidal radius being ill-defined (see \paperi{} for discussion), we have demonstrated in \paperii{} that none of this matters; the simulation outcome is insensitive to whether we truncate the subhalo at the virial radius or at the initial tidal radius.

Throughout we adopt model units in which the gravitational constant, $G$, the initial virial radius of the subhalo, $\rsub$, and the initial virial mass of the subhalo, $\Msub$, are all unity. With this choice, the initial virial velocity of the subhalo, $\Vsub$ is unity, while the host halo has $\Vhost = \calM^{1/3}$. Both have the same crossing time, $t_{\rm cross} \equiv \rvir/\Vvir = 1$. In physical units, the crossing time is
\begin{equation}\label{tcross}
t_{\rm cross} = 0.978 h^{-1} \Gyr \, \left(\frac{\Delta_{\rm vir}}{200}\right)^{-1/2}\,.
\end{equation}
where $h = H_0/(100 \kmsmpc)$. In what follows, whenever we quote time scales in physical units, we adopt $h=0.678$ \citep{Planck2016} and $\Delta\sub{vir} = 200$, which implies that a time interval of $(\Delta t)_{\rm model}=1$ corresponds to $1.44 \Gyr$. Note, though, that one can scale all the physical time scales quoted in this paper to other values of $H_0$ and $\Delta_{\rm vir}$ by simply multiplying the values quoted by the factor
\begin{equation}\label{gammascaling}
\Gamma(h, \Delta\sub{vir}) \equiv \left(\frac{h}{0.678}\right)^{-1}
 \, \left(\frac{\Delta\sub{vir}}{200}\right)^{-1/2}\,.
\end{equation}
Put differently, the mapping between time scales in DASH model units and physical units, is given by: $(\Delta t)_{\rm physical} = 1.44 \Gyr \, \Gamma(h, \Delta\sub{vir}) \, (\Delta t)_{\rm model}$.

\begin{figure}
\includegraphics[width=0.45\textwidth]{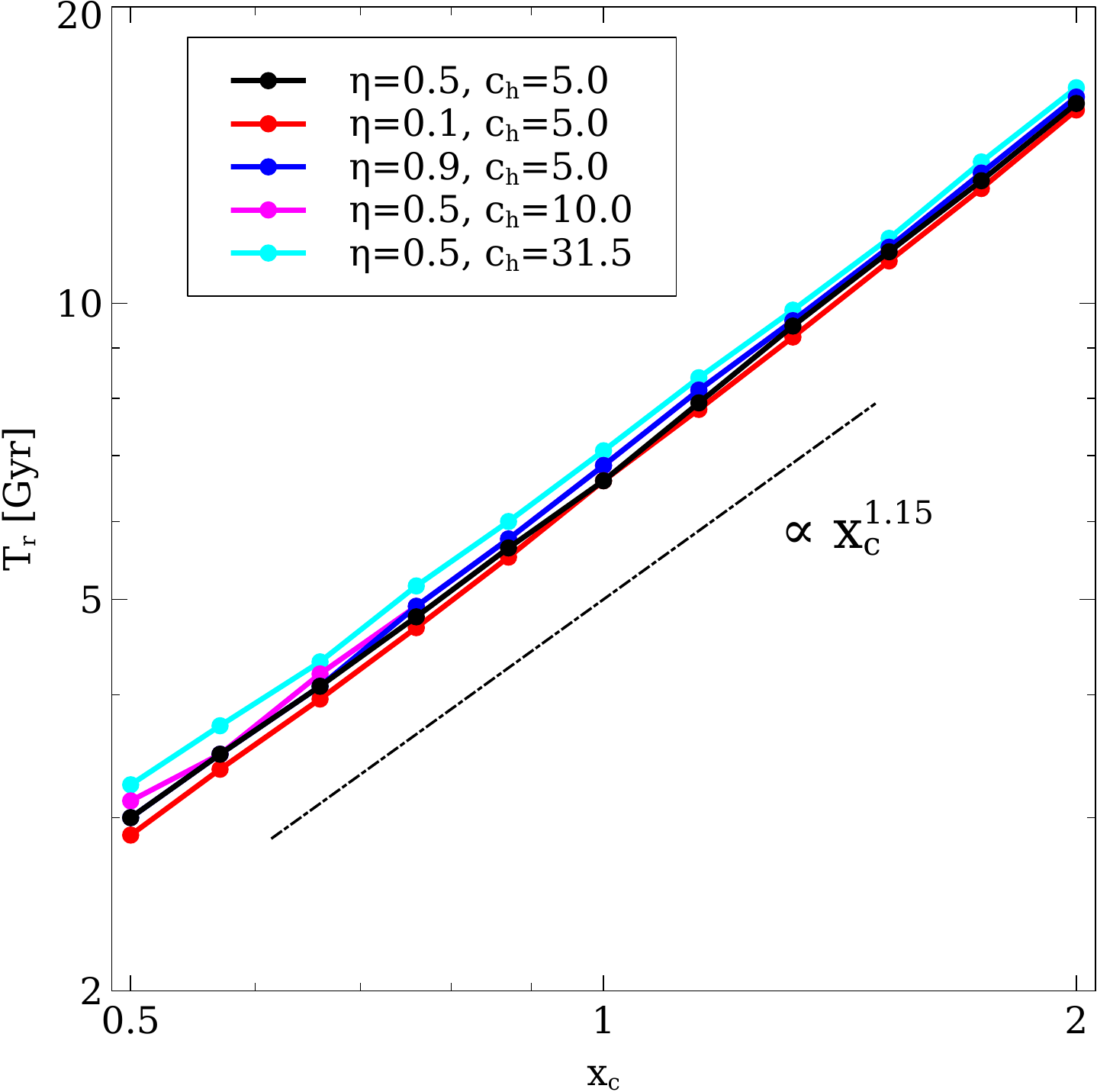}
\caption{The radial period, $T\sub{r}$, as a function of the parameter that controls the orbital energy of the subhalo, $x\sub{c}$. A virial overdensity of $\Delta\sub{vir}=200$ and redshift of $z=0$ are assumed. $T\sub{r}$ scales as $T\sub{r} \propto x\sub{c}^{1.15}$, and it is almost independent of $\eta$ and $c\sub{h}$.}
\label{fig:xc_tr}
\end{figure}

The DASH simulations span a wide range in orbital energy, $E$, and angular momentum, $L$. For convenience, we characterize the orbits using the following two dimensionless quantities:
\begin{itemize}
\item $x_\rmc \equiv r_\rmc(E)/\rhost$, the radius of the circular orbit corresponding to the orbital energy, $E$, expressed in terms of the virial radius of the host halo.
\item $\eta \equiv L/L_\rmc(E)$, the orbital circularity, defined as the ratio of the orbital angular momentum, $L$, and the angular momentum $L_\rmc(E)$ corresponding to a circular orbit of energy, $E$. Radial and circular orbits have $\eta = 0$ and 1, respectively.
\end{itemize}
We initially position the subhalo at the apocentre of its orbit, and follow its dynamical evolution for a period of $T_{\rm sim} = 36 \Gyr$. The orbit's radial period is given by
\begin{equation}\label{eq:tr}
T\sub{r} = 2 \int^{r\sub{a}}_{r\sub{p}} \frac{\rmd r}{\sqrt{2 [E - \Phi\sub{h}(r)] - L^2/r^2}}\,,
\end{equation}
\citep[e.g.,][]{Binney2008}, with $r\sub{p}$ and $r\sub{a}$ the pericentric and apocentric radii of the orbit, respectively, and $\Phi\sub{h}(r)$ the gravitational potential due to the host halo. The latter is given by
\begin{equation}
\Phi\sub{h}(r) = -\Vhost^2 \frac{\ln(1+\chost x)}{f(\chost) \, x}
\end{equation}
where $x = r/\rhost$ and 
\begin{equation}\label{fc}
f(c) = \ln(1+c) - c/(1+c)\,.
\end{equation}
Fig.~\ref{fig:xc_tr} plots $T\sub{r}$ as function of $x\sub{c}$ for several values of $\eta$ and $c\sub{h}$. Note that, to good approximation, $T\sub{r} \sim 6.7\Gyr \, x_\rmc^{1.15}$, with only a very weak dependence on $\eta$ or $c\sub{h}$. Hence, for the range of orbits covered by DASH, which have $x_\rmc \in [0.5,2.0]$ (see \S\ref{ssec:orb_c_params} below), $T_{\rm sim} = 36\Gyr$ corresponds to between 2.5 and 12 radial periods. The initial (Cartesian) vectors of the subhalo position and velocity with respect to the host halo are given by $\bX = (r_\rma, 0, 0)$ and $\bV = (0, L/r_\rma, 0)$, respectively, such that the subhalo orbit is confined to the $x-y$ plane. For each simulation we output a total of 301 snapshots, with the time interval between snapshots fixed at 0.12\,Gyr.

%%%%%%%%%%%%%%%%%%%%%%%%%%%%%%%%%%%%%%%
\subsubsection{Numerical techniques}
\label{sssec:numtech}

All DASH simulations have been carried out using a tree code \citep{Barnes1986} specifically developed for graphics processing unit (GPU) clusters \citep{Ogiya2013}. The code uses CPU cores to construct octree structures of the $N$-body particles, while GPU cards are used to compute gravitational accelerations through tree traversal. We employ 1,048,576 particles in each simulation. Forces between particles are softened using a Plummer softening length $\epsilon = 0.0003 \rsub$ (see \paperii{}) and the opening angle of the tree algorithm is set to $\theta=0.7$. Orbit integration uses a leapfrog scheme with a global, adaptive time step $\Delta t = 0.2 \sqrt{\epsilon/a\sub{max}}$ \citep[][]{Power2003}. Here $a\sub{max}$ is the maximum, absolute value of acceleration among all particles at that time\footnote{Typically the time step is fairly constant throughout the entire simulation, except for a small (typically factor $\sim 2$) decrease in $\Delta t$ during a high-speed pericentric passage of the host halo.}.  As we have demonstrated in \paperii{}, these numerical parameters are adequate to properly resolve the tidal evolution of subhaloes. In order to verify this, we have run a subset of our simulations at ten times better mass resolution (using $N \sim 10^7$ particles), which yields results that are indistinguishable from our nominal mass resolution.

%%%%%%%%%%%%%%%%%%%%%%%%%%%%%%%%%%%%%%%
\subsubsection{Data analysis and products}
\label{sssec:analysis}

For each simulation output we compute the bound mass fraction of the subhalo, $\fbound(t)$, using the iterative method described in detail in Appendix~A of \paperi{}. Briefly, the centre-of-mass position, $\br_{\rm com}$, and velocity, $\bv_{\rm com}$, are computed using the five percent most bound particles. These quantities are subsequently used to compute the binding energy of each particle.  This is iterated until the changes in $\br_{\rm com}$ and $\bv_{\rm com}$ are smaller than $10^{-4} \rsub$ and $10^{-4} \Vsub$, respectively, which typically requires 3-10 iterations. The bound fraction, $\fbound(t)$, is then defined as the fraction of the original particles that at time $t$ have a negative binding energy.

All simulations in the DASH library initially have sufficient numerical resolution to properly resolve the dynamics of the dark matter subhalo. However, as highlighted in \paperii{}, simulation results can become unreliable once the subhalo has lost a significant fraction of its initial mass due to tidal evolution. \paperii{} presented the two criteria that a subhalo in a numerical simulation needs to satisfy in order for its dynamical evolution to be  reliable. The first criterion, which is motivated by the work of \citet{Power2003}, tests whether the softening length is sufficiently small to resolve the relevant accelerations. It translates into a requirement for the bound mass fraction given by
\begin{equation}
\fbound(t) > 1.79 \, \frac{\csub^2}{f(\csub)} \, \left(\frac{\epsilon}{\rsub}\right) \, \left(\frac{r_\rmh(t)}{\rsub}\right)
\end{equation}
with $r_\rmh(t)$ the half-mass radius of the (bound part of) the subhalo at time $t$. Using that all DASH simulations adopt $\varepsilon = 0.0003 \rsub$, we thus have that the simulation results need to satisfy
\begin{equation}\label{critA}
\fbound(t) > \frac{0.054}{f(\csub)} \, \left(\frac{\csub}{10}\right)^2 \, \left(\frac{r_\rmh(t)}{\rsub}\right)\,, 
\end{equation}
for the results to be deemed reliable.  The second criterion is related to discreteness noise, and puts a constraint on the number of bound particles in the subhalo. In particular, it states that the number of bound particles needs to exceed 
\begin{equation} \label{Ncrit}
N\sub{crit} \equiv 80 N^{0.2}\,.
\end{equation}
with $N$ the number of particles in the {\it initial} subhalo. Once the number of bound particles falls below this critical value, the subhalo starts to experience a runaway instability, triggered by discreteness noise, which quickly leads to its demise (i.e., artificial disruption). Since all DASH simulations have $N = 1,048,576$, we have that $N\sub{crit} = 1,280$, which implies that the DASH simulations are only reliable for 
\begin{equation}\label{critB}
\fbound(t) > 1.22 \times 10^{-3}\,.
\end{equation}
As discussed in \S\ref{ssec:data_str} below, the DASH library contains, for each output, the bound mass fraction and the half-mass radius of the subhalo, which the user can use to test whether the output satisfies both criteria. More than 99.5\% of all the simulation outputs available in the DASH library satisfy both criteria~(\ref{critA}) and~(\ref{critB}).

%%%%%%%%%%%%%%%%%%%%%%%%%%%%%%%%%%%%%%%%%%%%%%%%%%%%%%%%%%%%%%%%%%%%%%%%%%%%%%%%%%%%
\subsection{The DASH parameter space}
\label{ssec:orb_c_params}

The simulations described above are characterized by six parameters: the mass and concentration of the host halo, $\Mhost$ and $\chost$, respectively, the mass and concentration of the subhalo, $\Msub$ and $\csub$, respectively, and the orbital parameters $\xc$ and $\eta$. In order to limit the numerical cost of sampling this six-dimensional parameter space without sacrificing the volume sampled, we adopt a strategy that devotes more computational efforts to the regions of parameter space with higher probability. In doing so, we are guided by physical considerations and results from cosmological simulations.

%%%%%%%%%%%%%%%%%%%%%%%%%%%%%%%%%%%%%%%
\subsubsection{Initial mass ratio}
\label{sssec:calm}

Since dynamical timescales and the impact of tidal forces depend only on density, and since all haloes, by definition, have the same virial density, our simulation results should be independent of $\Mhost$ and $\Msub$, significantly reducing the dimensionality of our parameter space. However, there is one important caveat. In reality, a subhalo orbiting a host halo experiences a dynamical drag force caused by the (perturbed) matter in the host halo \citep[`dynamical friction'; e.g.,][]{Chandrasekhar1943} and by its own stripped material \citep[`self-friction'; e.g.,][]{Fujii2006,Fellhauer2007}. The impact of these drag forces, which result in orbital decay, increase strongly with decreasing mass ratio $\calM = \Mhost/\Msub$. When $\calM \gta 100$, though, their impact can safely be ignored \citep[e.g.,][]{Mo2010}. Throughout, we therefore restrict ourselves to simulations with $\calM = 1000$. As we specifically demonstrate in \S\ref{ssec:mass_ratio}, these simulations accurately capture the tidal evolution of subhaloes for any $\calM \gta 100$, independent of the absolute value of $\Mhost$.

%%%%%%%%%%%%%%%%%%%%%%%%%%%%%%%%%%%%%%%
\subsubsection{Sampling orbital parameters}
\label{sssec:orb}
\begin{figure}
\includegraphics[width=0.45\textwidth]{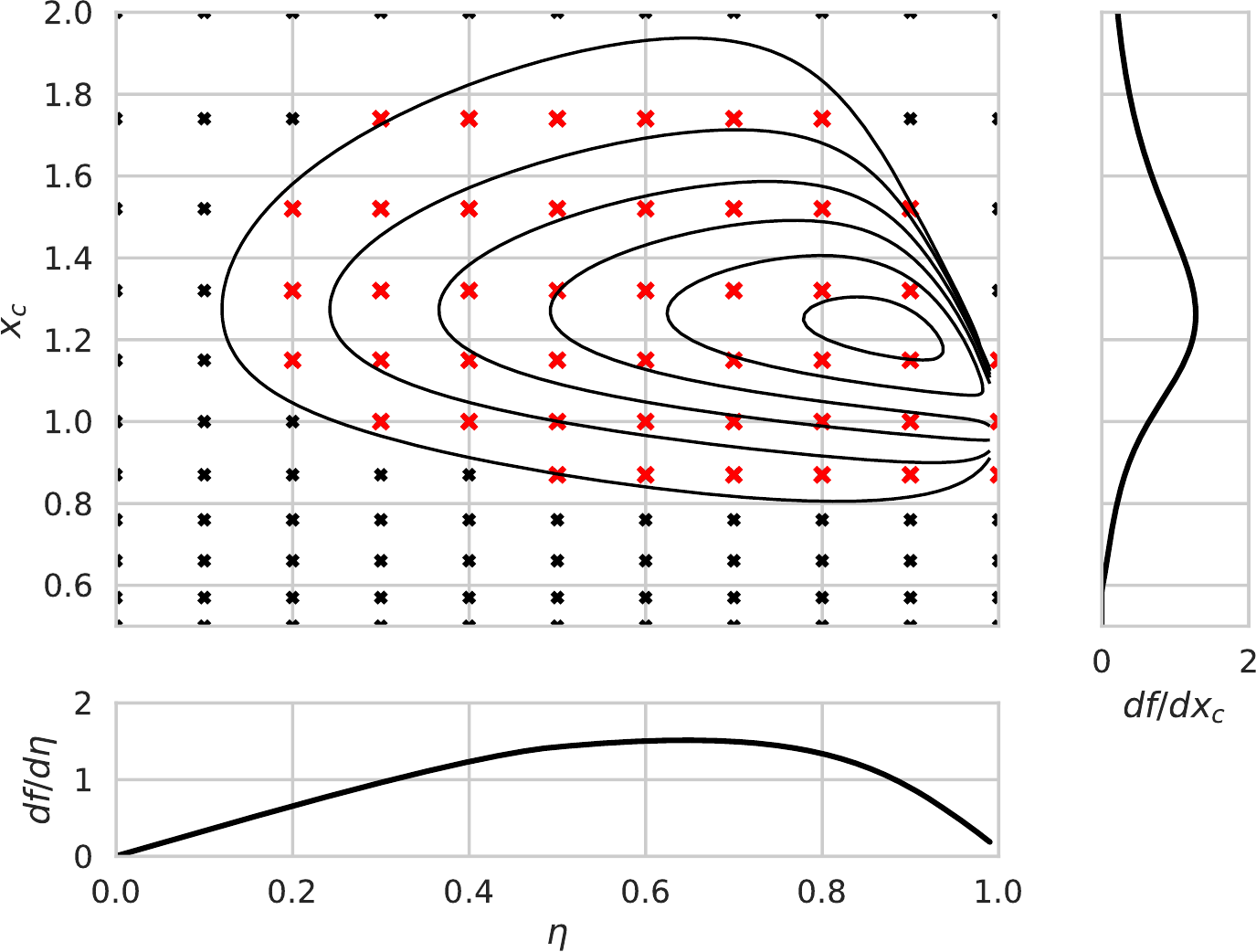}
\caption{Probability distribution of the orbital parameters, $x\sub{c}$ and $\eta$, derived from the fitting function by \citet{Jiang2015} for a host halo mass of $M=10^{12}$\,\msun{} and minor merger mass ratios in the range $0.0001 \leq M\sub{s}/M\sub{h} \leq 0.005$. The host halo is assumed to follow a NFW density profile with a concentration $c_\rmh = 5$. Crosses represent the parameter sets used in the DASH library. Red crosses cover the regions of parameter space with the largest probabilities in the distribution and most simulations are devoted to these regions. The right and bottom subset panels show the one dimensional marginalized probability distributions of $x\sub{c}$ and $\eta$, respectively. 
\label{fig:dist_orb_contour}}
\end{figure} 

Using a state-of-the-art cosmological simulation, \cite{Jiang2015} studied the orbital parameters of dark matter subhaloes at their moment of accretion into their host halo \citep[see also e.g.,][]{Tormen1997, Zentner2005, Khochfar2006, Wetzel2011, vandenBosch2017}. In particular, \cite{Jiang2015} measured the radial and tangential components, $V\sub{r}$ and $V\sub{\theta}$, of the relative velocity vector between the host- and subhaloes at infall.  We convert their bivariate distribution of $V\sub{r}/\Vhost$ and $V\sub{\theta}/\Vhost$ to the corresponding bivariate distribution of $x_\rmc$ and $\eta$, assuming an NFW density profile for the host halo with a concentration parameter $c\sub{h}=5$. The resulting PDF $P(x_\rmc,\eta)$ is shown in Fig.~\ref{fig:dist_orb_contour}, where the small panels show the corresponding, marginalized distributions of $x_\rmc$ (side panel) and $\eta$ (bottom panel).

In order to ensure that our sampling of $x_\rmc$ and $\eta$ covers the entire range relevant for modelling the assembly and evolution of dark matter substructure, we proceed as follows.  We sample $\eta$ linearly over the entire range from $\eta=0$ (purely radial orbit) to $1.0$ (circular orbit) in 10 steps of $\Delta\eta= 0.1$, and $x_\rmc$ logarithmically over the range from $x_\rmc=0.5$ to $2.0$, in 10 steps of $\Delta\log x_\rmc \simeq 0.06$. Hence, there are a total of 121 combinations of $(x_\rmc,\eta)$. Together with 121 combinations of $(\chost,\csub)$ (see below), this would imply that we need to run more than 14,500 simulations. In addition to being prohibitively expensive, this is also unwarranted, as we will demonstrate in \S\ref{sec:ml}. Instead, we significantly reduce the number of simulations by focusing most of our efforts on the more likely parts of parameter space. The red crosses in Fig.~\ref{fig:dist_orb_contour} indicate the 45 most likely combination of $x_\rmc$ and $\eta$. For each of these we run the 45 most likely combinations of $(\chost,\csub)$, using the sampling strategy detailed below. The black crosses, on the other hand, indicate the combinations of $(x_\rmc,\eta)$ for which we only run one realization of halo concentrations, namely the one for $(\chost,\csub) = (5.0,10.0)$.

%%%%%%%%%%%%%%%%%%%%%%%%%%%%%%%%%%%%%%%
\subsubsection{Sampling halo concentrations}
\label{sssec:halo_c}
\begin{figure}
\includegraphics[width=0.45\textwidth]{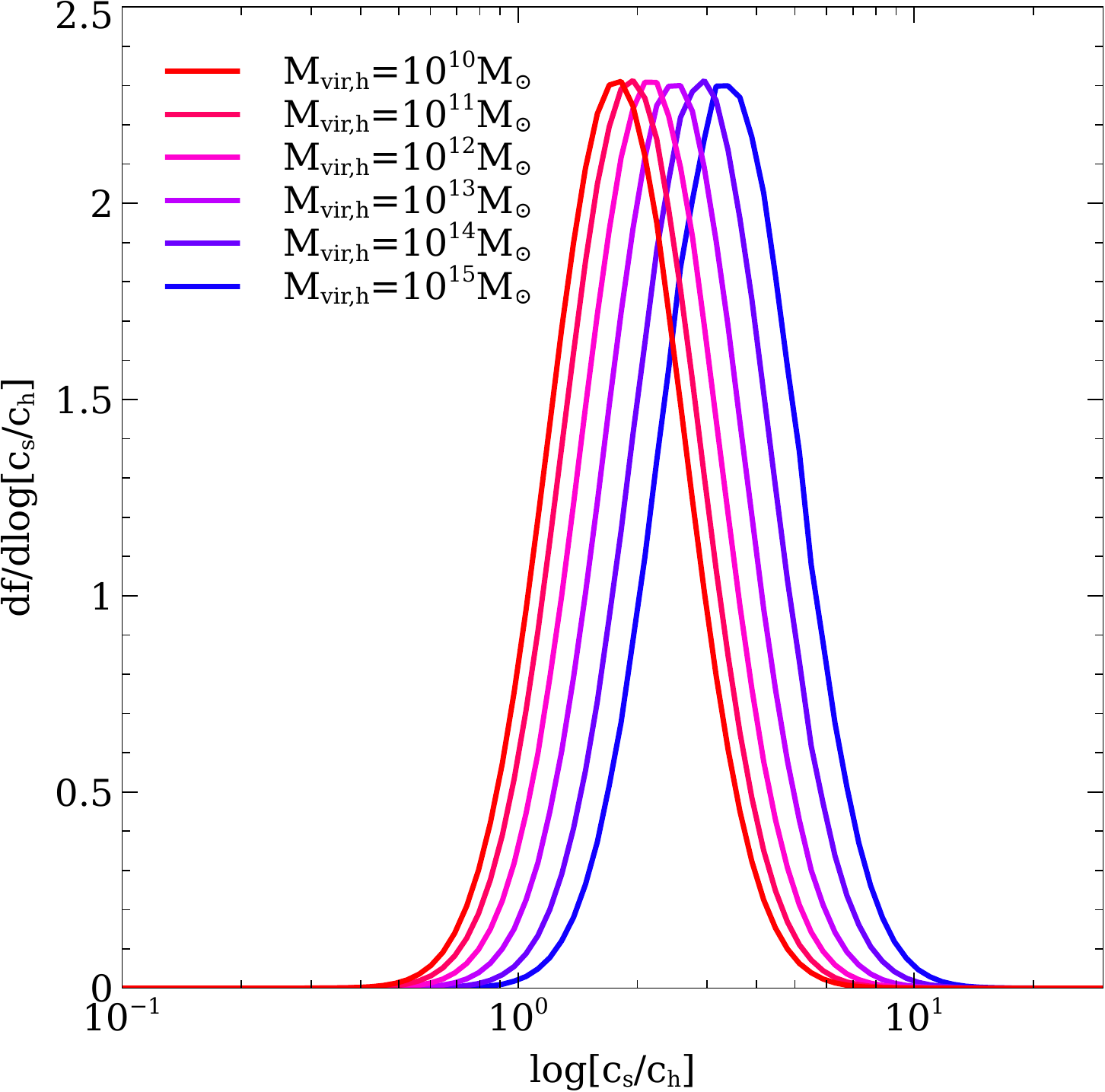}
\caption{Probability distribution of the halo concentration ratio, $c\sub{s}/c\sub{h}$, based on the halo merger rate by \citet{Fakhouri2010} and $c(M\sub{vir},z)$ relation by \citet{Ludlow2016}, as well as the log-normal scatter of $\sigma_{\log c} = 0.12$ in the $c(M\sub{vir},z)$ relation. A redshift of $z=0$ is assumed. Lines represent the distribution for host haloes with $M\sub{vir,h} = 10^{10}, 10^{11}, 10^{12}, 10^{13}, 10^{14}$ and $10^{15}$\,\msun{}, respectively.
\label{fig:fcratio}}
\end{figure}

\begin{figure}
\includegraphics[width=0.45\textwidth]{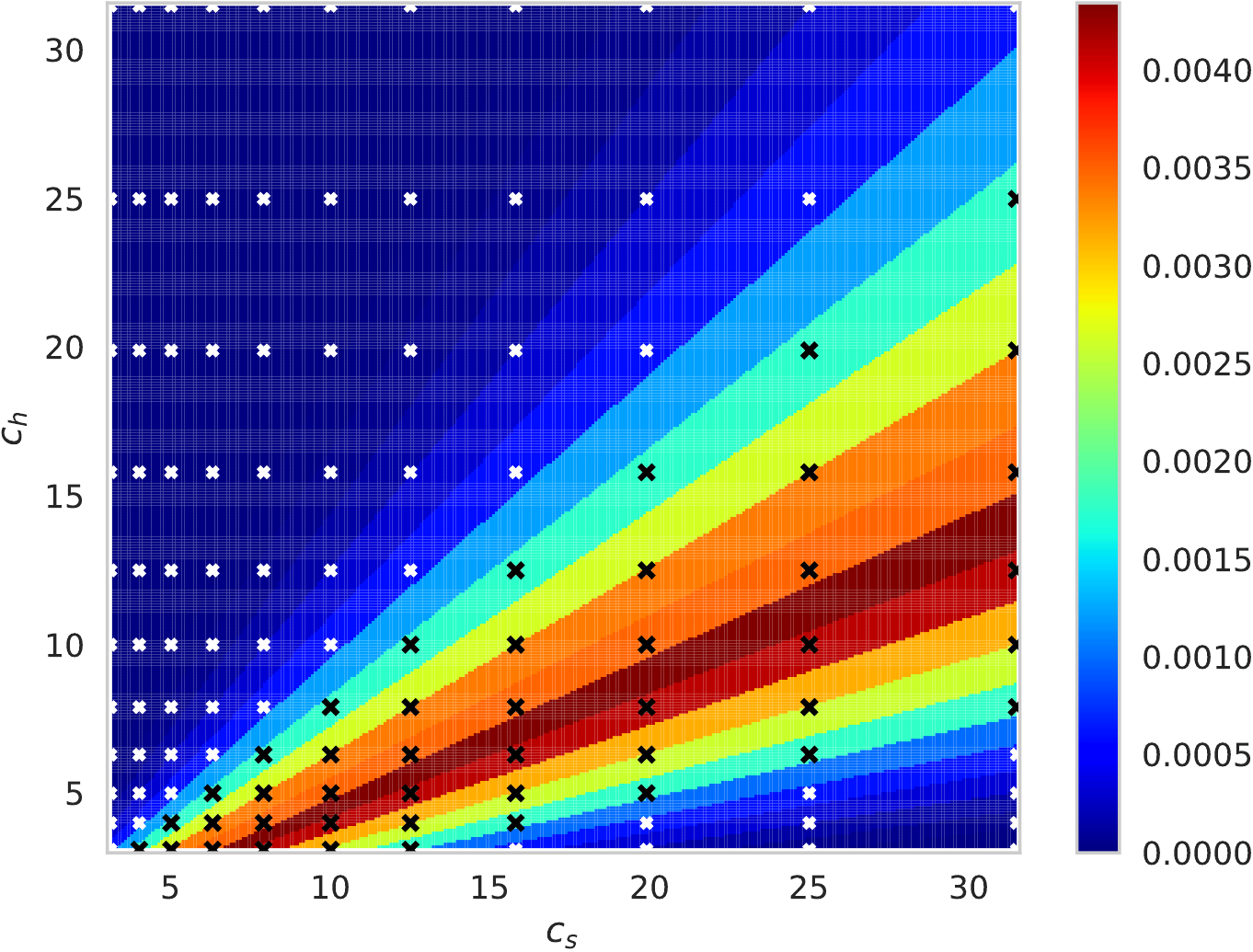}
\caption{ Probability distribution of the halo concentration parameters, $c\sub{h}$ and $c\sub{s}$. The line for $M\sub{vir,h}=10^{12}$\,\msun{} in \autoref{fig:fcratio} is translated into this two-dimensional space, assuming that the distribution in \autoref{fig:fcratio} is independent of $M\sub{vir,h}$. Crosses represent the parameter sets studied in the DASH library. Black crosses cover the regions of largest probability in the distribution and most simulations are devoted to these concentration pairs.
\label{fig:dist_str}}
\end{figure}

In deciding on how to sample $\chost$ and $\csub$, we are again guided by cosmological simulations. These reveal that haloes of fixed mass have a log-normal distribution of halo concentrations, with a scatter of $\sim 0.12$\,dex, and with a median that decreases with halo mass roughly as $c \propto M^{-0.1}$ \citep[e.g.,][]{Bullock2001,  Prada2012, Dutton2014, Diemer2015, Ludlow2016}. For DASH we sample both $\csub$ and $\chost$ using $\log c = 0.5, 0.6, 0.7,...,1.5$.  This covers the range $3.1 \lta \cvir \lta 31.5$, which is adequate for the vast majority of all haloes in our mass range of interest (roughly $10^7 < \Mvir/(\Msunh) < 10^{15}$).  This sampling, though, yields a total of 121 unique pairs of $(\chost,\csub)$. Similar to our sampling of the orbital parameters, we focus most of our efforts on the `most likely' combinations of $\chost$ and $\csub$. First, using the fitting function of \cite{Fakhouri2010}, we compute the halo merger rate as a function of the mass ratio of the merging haloes. We only consider minor mergers with a mass ratio $\calM > 100$, representative of our simulations. Next, we draw concentrations for each of the merging haloes using the concentration-mass relation of \citet{Ludlow2016}, and account for log-normal scatter with $\sigma_{\log c} = 0.12$. Using this Monte-Carlo procedure, we compute the PDF for $\log[\csub/\chost]$. Results for different values of $\Mhost$ are shown in Fig.~\ref{fig:fcratio}, and are in good agreement with the results obtained from cosmological simulations (e.g., \paperi{}). Note that more massive host haloes tend to have somewhat higher concentration ratios on average. However, the strength of this mass dependence is weak compared to the width of the individual PDFs, and we therefore ignore it in what follows. Fig.~\ref{fig:dist_str} shows the joint distribution of $c\sub{h}$ and $c\sub{s}$, derived using the PDF for $\log[\csub/\chost]$ for a host halo of mass $\Mhost = 10^{12} \Msunh$ and adopting a uniform distribution\footnote{Although halo concentrations follow a log-normal distribution at fixed mass, we adopt (for simplicity) a uniform distribution such that the simulations are relevant for a wide range in halo masses.} in $\chost$. Black crosses indicate combinations of $(\chost,\csub)$ for which we have run simulations for all 45 most likely orbital parameter combinations $(x_\rmc,\eta)$ (the red crosses in Fig.~\ref{fig:dist_orb_contour}). White crosses, on the other hand, indicate combinations of $(\chost,\csub)$ for which we have run only one orbital configuration, namely with $(x_\rmc,\eta) = (1.0,0.5)$.

%%%%%%%%%%%%%%%%%%%%%%%%%%%%%%%%%%%%%%%%%%%%%%%%%%%%%%%%%%%%%%%%%%%%%%%%%%%%%%%%%%%%
\subsection{Data structure}
\label{ssec:data_str}

Using the sampling strategy outlined above, there are a total of 2,177 idealized $N$-body simulations that we have run. It is hierarchically structured, with different layers of directories corresponding to the various adopted orbital or concentration parameters. Each directory presents the results of the simulation in the form of five simple text files (see \autoref{tab:files}), and each file contains a header listing the parameter set used in the simulation, as well as a brief description of the data file. Rather than accessing the results for individual simulations, the user can also download a single compressed archive file ($\sim 0.6$\,GB) containing the results for all 2,177 simulations as well as a Python routine that uses a non-parametric model based on random forest regression (see \S\ref{sec:ml}), to predict the bound mass fraction of a subhalo as a function of time, for any configuration captured by our parameter space. In what follows we briefly describe the information that is provided for each simulation.
\begin{table*}
  \caption{Summary of the DASH data files available for each simulation \label{tab:files}}
\begin{tabular}{cc}
\hline 
filename       & description\\
\hline
\texttt{subhalo\_evo}      & position, velocity, bound mass fraction, and half-mass radius of subhalo as function of $t$ \\
\texttt{radprof\_rho}     & subhalo density, $\rho(r,t)$, as function of $r$ and $t$, in units of $200\rho\sub{crit}(0)$ \\
\texttt{radprof\_m}        & enclosed mass of subhalo, $M(<r,t)$, as function of $r$ and $t$, in units of $M\sub{vir,s}$ \\
\texttt{radprof\_sigmar}  & radial velocity dispersion, $\sigma\sub{r}(r,t)$, as function of $r$ and $t$, in units of $\Vsub$ \\
\texttt{radprof\_sigmat}  & tangential velocity dispersion, $\sigma\sub{t}(r,t)$, as function of $r$ and $t$, in units of $\Vsub$ \\
\hline
\end{tabular}
\end{table*}

The file \texttt{subhalo\_evo} lists the temporal evolution of several subhalo properties. The first column lists the ID of each simulation snapshot, which can be used to compute the corresponding physical time as
\begin{equation} \label{eq:phys_time}
t = 0.12 \Gyr \, \Gamma(h, \Delta\sub{vir}) \times {\rm ID}\,.
\end{equation}
Columns 2-4 and 5-7 list the Cartesian components of the position and velocity vectors of the centre-of-mass of the subhalo, respectively. The former and latter are normalized by the virial radius and velocity of the host halo, $\rhost$ and $\Vhost$, respectively. Finally, columns 8 and 9 list the subhalo bound mass fraction, $\fbound(t)$, and the subhalo's half-mass radius, $r_\rmh(t)$.  The latter is normalized by the virial radius of the initial subhalo, $\rsub$. Note that these outputs describe the tidal evolution of a (spherical NFW) subhalo of any mass, $\Msub$, in a (spherical NFW) host halo of any mass, $\Mhost$, as long as $\calM = \Mhost/\Msub \gta 100$. For smaller mass ratios dynamical friction, which is not accounted for, is important.

The other data available for each simulation in the DASH library are four text files that list the time evolution of four different radial profiles: the normalized density, $\rho(r,t)/[200\,\rho_{\rm crit}(0)]$, the enclosed, normalized mass profile, $M(<r,t)/\Msub$, and the normalized radial and tangential velocity dispersion profiles, $\sigma\sub{r}(r,t)/\Vsub$ and $\sigma\sub{t}(r,t)/\Vsub$, respectively. The positions and velocities of all particles are defined with respect to the centre-of-mass position and velocity of the subhalo, and only bound particles are taken into account.  The first row of each of the \texttt{radprof} files lists the radial bins, which span the range $-2.95 \leq \log(r/\rsub) \leq 0.95$ in equally spaced bins of width $\Delta\log(r/\rsub) = 0.1$. The subsequent rows list the radial profiles for each of the 301 simulation snapshots, with row $j$ corresponding to snapshot ID=$j-2$. The corresponding physical time is given by equation~(\ref{eq:phys_time}).

%%%%%%%%%%%%%%%%%%%%%%%%%%%%%%%%%%%%%%%%%%%%%%%%%%%%%%%%%%%%%%%%%%%%%%%%%%%%%%%%%%%%
%%%%%%%%%%%%%%%%%%%%%%%%%%%%%%%%%%%%%%%%%%%%%%%%%%%%%%%%%%%%%%%%%%%%%%%%%%%%%%%%%%%%
\section{Some examples from the simulations}
\label{sec:examples}

While classical dynamical friction is absent in the simulations, another source of friction caused by the mass stripped from the subhalo may alter the mass and orbital evolution of the subhalo in the tidal interactions. \S\ref{ssec:mass_ratio} studies the validity condition that must be satisfied in order to neglect this friction force. The subsequent subsections present some examples of the DASH simulations, namely the mass evolution (\S\ref{ssec:fb}) and radial profiles (\S\ref{ssec:profs}) of the tidally stripped subhaloes.

%%%%%%%%%%%%%%%%%%%%%%%%%%%%%%%%%%%%%%%%%%%%%%%%%%%%%%%%%%%%%%%%%%%%%%%%%%%%%%%%%%%%
\subsection{Dependence on the initial mass ratio}
\label{ssec:mass_ratio}

\begin{figure*}
\begin{center}
\includegraphics[width=0.95\textwidth]{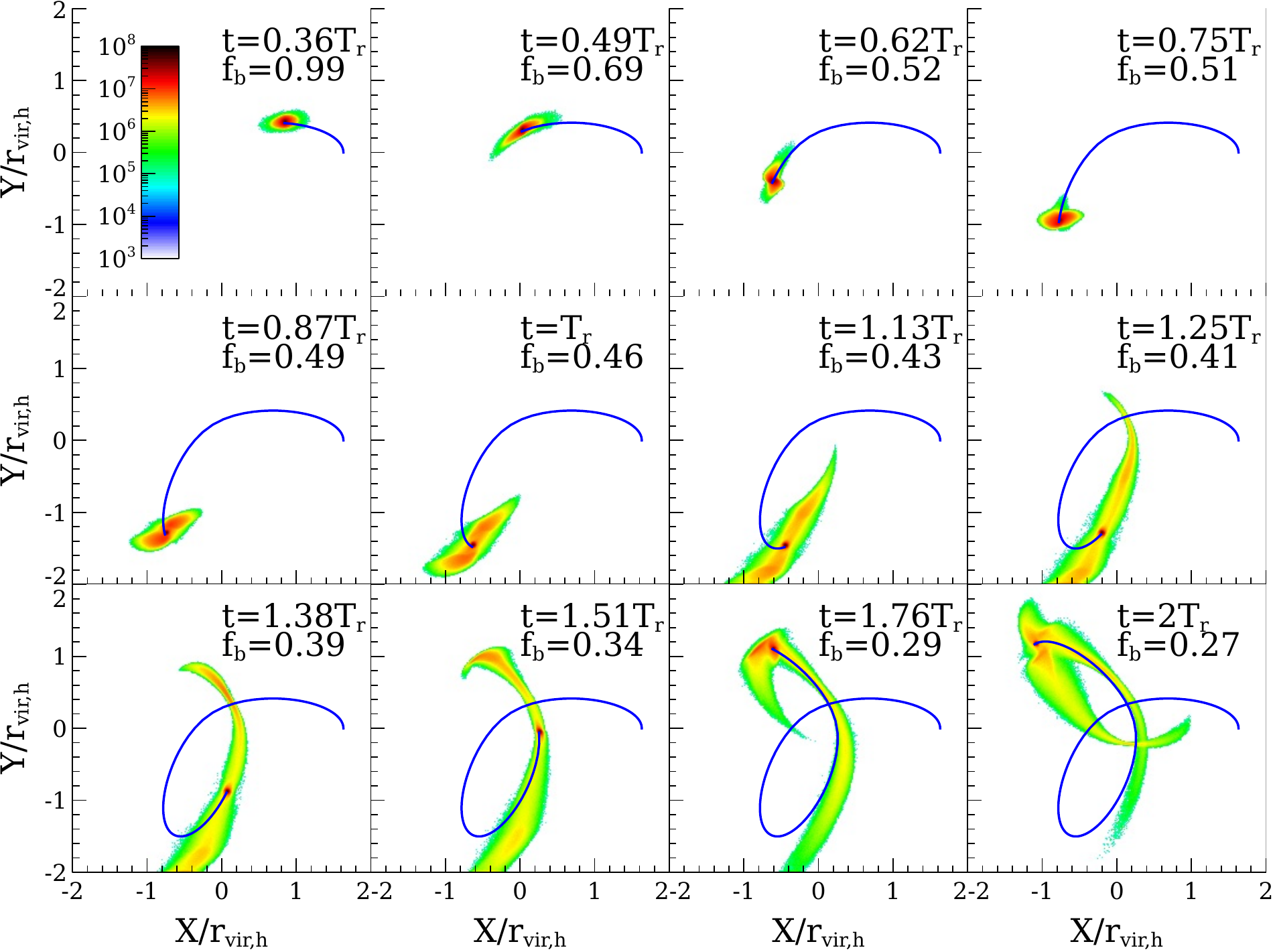}
\end{center}
\caption{
Distribution of particles in the simulation run with $x\sub{c}=1.00$, $\eta=0.5$, $c\sub{h}=5$, $c\sub{s}=10$ and $\calM=10^3$. The blue line represents the subhalo orbit with respect to the centre of the host halo, which corresponds to the origin in this plot. From the leftmost panel in the top row to the rightmost panel in the bottom row, the evolution is illustrated. The time and bound mass fraction, $f\sub{b}$, of each snapshot are denoted in each panel. The spatial coordinates are scaled by the virial radius of the host halo, $r\sub{vir,h}$, and the colour bar (column density) is given in arbitrary units. \label{fig:map_evolution}}
\end{figure*}

As mentioned in \S\ref{ssec:setup}, the host halo is modelled as an analytical NFW potential and the initial mass ratio between the host- and subhaloes, $\calM \equiv M\sub{vir,h}/M\sub{vir,s}$, is fixed at $10^3$ for all simulations. In this subsection, we discuss the validity condition for these assumptions.

In actual mergers, friction forces may alter the orbit of subhaloes. The classical dynamical friction \citep{Chandrasekhar1943} is caused by wakes formed in the density field of the larger host system due to the gravitational force of smaller objects, such as supermassive black holes, galactic bars, and dark matter subhaloes \citep[e.g.,][]{Weinberg2007,Antonini2012,Ogiya2016}. This kind of friction force is absent in our simulations because we employ an analytical host potential and the response in the density field of the host system due to the gravitational force of the subhalo is therefore not taken into account. The timescale of orbital decay due to this dynamical friction is roughly equal to $\calM \tau_{\rm ff}$ \citep[e.g.,][]{Mo2010}, where $\tau_{\rm ff}$ is the free-fall time of the host halo, which is of the order of 2-3\,Gyr at $z=0$. Since all our simulations adopt $\calM = 10^3$, is it clear that we can safely neglect the effects of dynamical friction, and thus use an analytical potential to model the host halo.

However, in addition to this `classical' dynamical friction, which is not accounted for in our simulations, there is an additional friction force that may contribute, and which {\it is} included in our simulations. This force is due to the mass stripped from the subhalo itself, and we therefore refer to this force as `self-friction'. \autoref{fig:map_evolution} shows an example of the tidal evolution of a subhalo in the DASH simulation with $x\sub{c}=1.0$, $\eta=0.5$, $c\sub{h}=5$, $c\sub{s}=10$ and $\calM=10^3$. Time progresses from the top left to the bottom right, as indicated, while colors indicate column density (in arbitrary units). As is evident, the tidal field of the host halo strips mass from the subhalo, giving rise to a leading and a trailing tidal arm that roughly trace out the subhalo's orbit. The gravitational force from this stripped material on the bound remnant, can result in a net deceleration, thus giving rise to a self-friction (\citealt{Fujii2006, Fellhauer2007}; \paperii{}). 

\begin{figure}
\begin{center}
\includegraphics[width=0.45\textwidth]{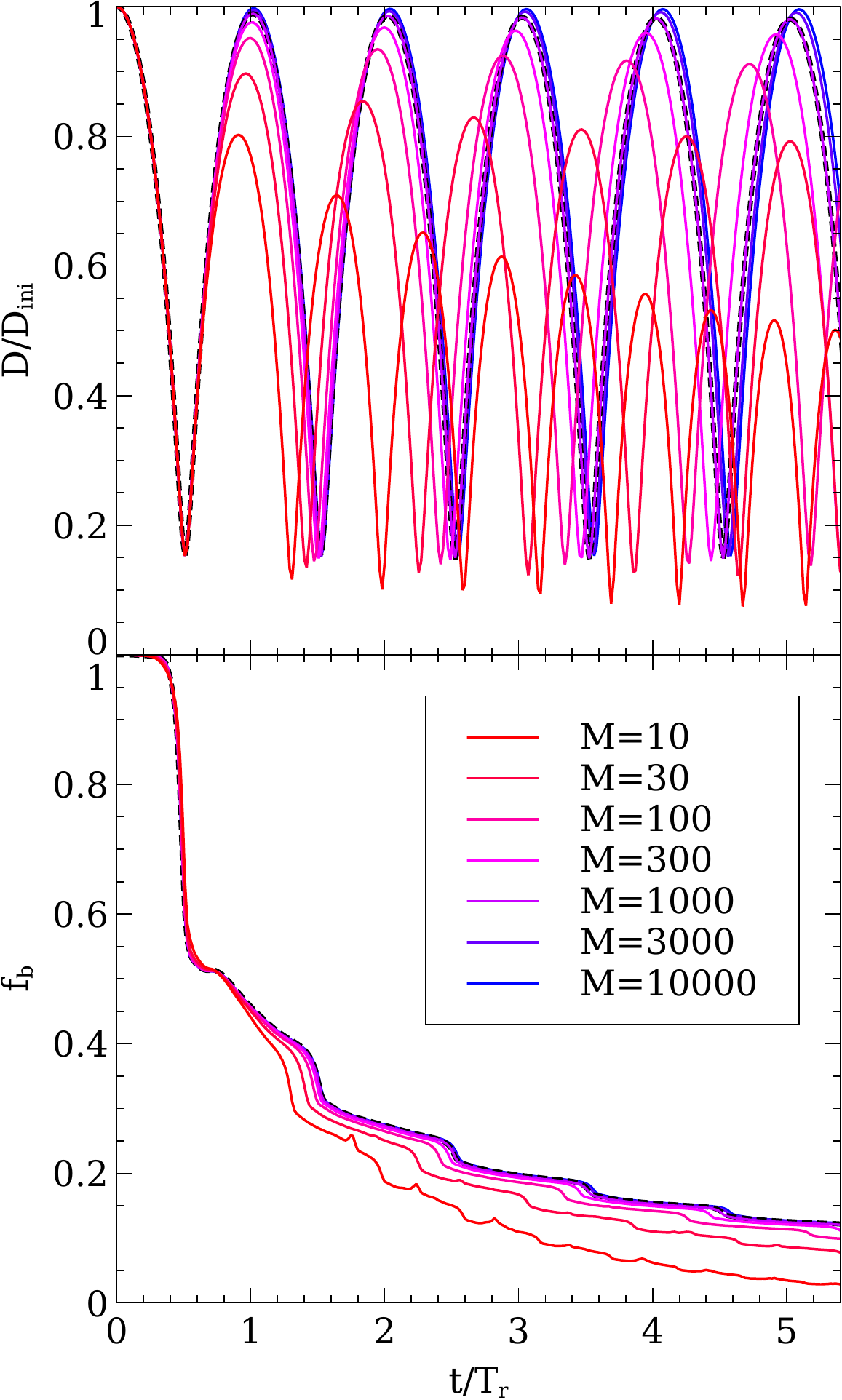}
\end{center}
\caption{
({\it Upper panel}) Distance between the centres of the host potential and subhalo, $D$, normalized by the initial value, $D\sub{ini}$. ({\it Lower panel}) Bound mass fraction, $f\sub{b}(t) \equiv M\sub{b}(t)/M\sub{vir,s}$, where $M\sub{b}(t)$ and $M\sub{vir,s}$ are the bound mass of the subhalo at each given time and the initial subhalo mass, respectively. The initial mass ratio between the host- and subhalo, $\calM \equiv M\sub{vir,h}/M\sub{vir,s}$, is denoted in the legend. In the simulations, the same parameters for the host potential ($M\sub{vir,h}$ is fixed and $c\sub{h}=5$) and subhalo orbit ($x\sub{c}=1.00$ and $\eta=0.5$) are adopted. While the concentration of the subhalo is also fixed ($c\sub{s}=10$), the subhalo mass is varied and its spatial scale is altered accordingly (i.e., $r\sub{vir,s} \propto M\sub{vir,s}^{1/3}$). In the simulations represented by solid lines, the standard resolution level ($N \approx 10^6$) is adopted. As an additional test, the number of particles is increased by a factor of ten in the simulation represented by the black dotted line ($\calM=10^3$), with other parameters held fixed. 
\label{fig:mass_ratio}}
\end{figure}

In order to estimate the impact of self-friction, we have performed a number of simulations in which we vary the mass ratio, $\calM$, while keeping all other parameters fixed to $x\sub{c}=1.0$, $\eta=0.5$, $c\sub{h}=5$ and $c\sub{s}=10$. Each subhalo is simulated with 1,048,576 particles and uses our fiducial softening and time-stepping. The results are shown in \autoref{fig:mass_ratio}, which plots the orbital radius (upper panel) and bound mass fraction (lower panel) as functions of time. The black dotted line corresponds to the results from a $\calM=10^3$ simulation with an order of magnitude more particles ($N=10,485,760$), and is shown to demonstrate that these results are well converged; the black dotted line lies exactly on top of the results from our fiducial resolution simulation, shown as a purple solid line. In the absence of any friction, the results from all these simulations should all be identical. This follows from the scale-free nature of gravity, the fact that (sub)haloes with the same concentration parameter but different mass, have identical density profiles as function of the normalized radius $r/r_{\rm vir}$, and the fact that the tidal radius scales with the ratio of the densities of host and subhalo\footnote{A comprehensive review is found in \paperi{} (see also e.g., \citealt{Binney2008})}.

While the simulations with $\calM \gta 100$ are indeed indistinguishable, those with smaller mass ratios clearly deviate, both in terms of their orbit and in terms of the bound mass fraction. Notably, the apocentric distance reached after the first pericentric passage becomes smaller for smaller $\calM$, indicating that the subhalo is experiencing self-friction due to its own stripped material. Over time this self-friction causes the orbit to shrink, which reduces the pericentre of the orbit, thereby exposing the subhalo to a stronger tidal tidal, which in turn results in a reduced bound mass fraction. As is evident from  \autoref{fig:mass_ratio}, the impact of self-friction is more pronounced for smaller values of $\calM$. This is easy to understand; as $\calM$ decreases, the force from the stripped material on the bound remnant {\it relative} to that from the host halo increases, causing a more pronounced deceleration along its orbit. We present a more detailed study of self-friction in a forthcoming paper (Miller et al. 2019, in preparation). Relevant for this study, though, is that self-friction is negligible for $\calM \ga 100$, and that the evolution of the bound mass fraction does not depend on $\calM$ as long as this is the case. Hence, the DASH simulations presented here, which all have $\calM = 10^3$, are valid for any mass ratio  $\calM \ga 100$.

%%%%%%%%%%%%%%%%%%%%%%%%%%%%%%%%%%%%%%%%%%%%%%%%%%%%%%%%%%%%%%%%%%%%%%%%%%%%%%%%%%%%
\subsection{Bound mass fraction}
\label{ssec:fb}

\begin{figure*}
\begin{center}
\includegraphics[width=0.80\textwidth]{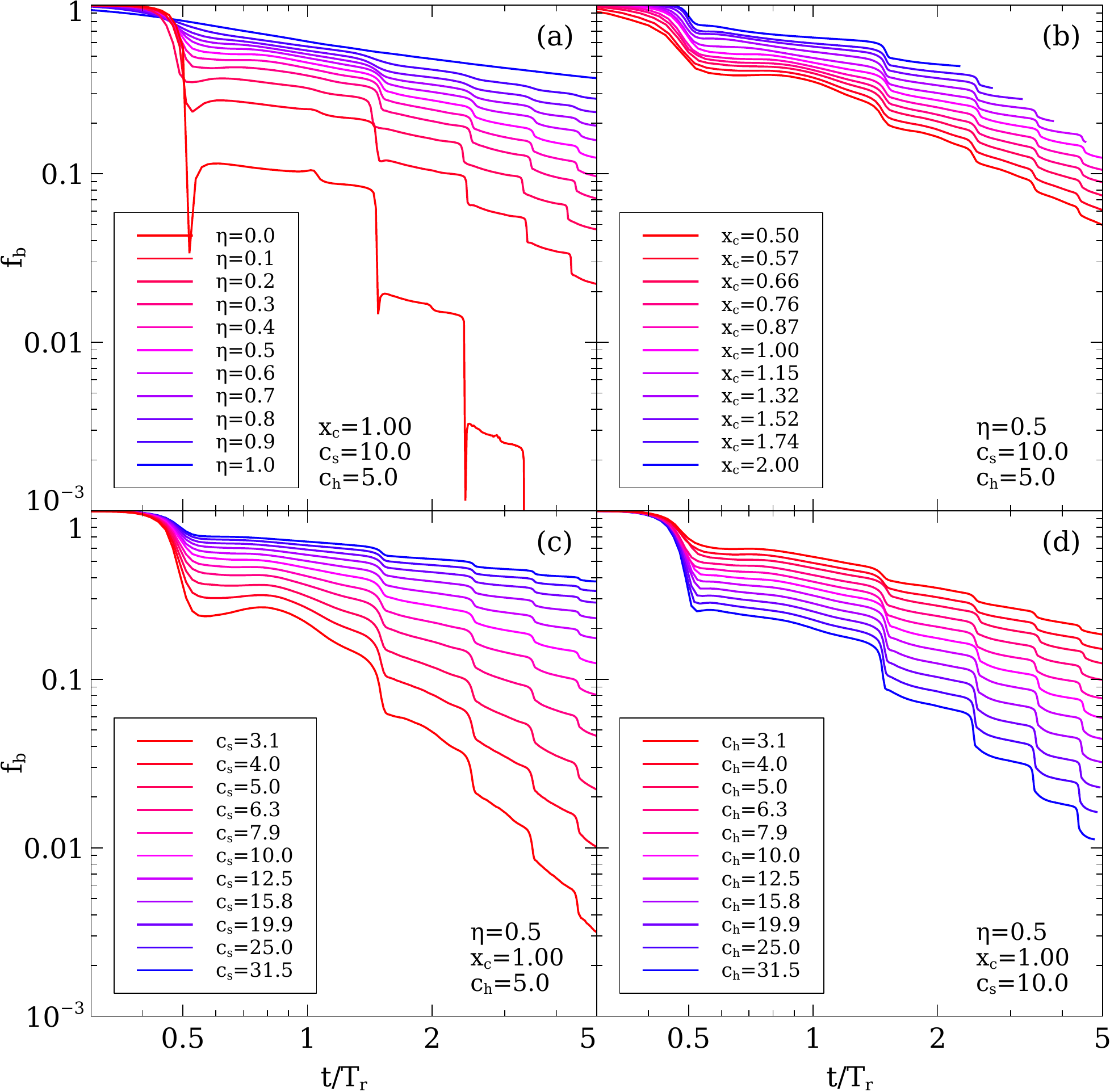}
\end{center}
\caption{
Evolution of the subhalo bound mass fraction, $f\sub{b}$. The timescale is normalized by the radial period, $T\sub{r}$, of each orbit. In panels (a), (b), (c), and (d), each of the parameters $\eta$, $x\sub{c}$, $c\sub{s}$, and $c\sub{h}$ are varied, respectively, holding the others fixed as denoted. 
\label{fig:fb}}
\end{figure*}

\autoref{fig:fb} demonstrates the evolution of the bound mass fraction of the subhalo, $f\sub{b}$, obtained in some example simulations. Fixing the structural parameters of the host- and subhaloes (upper panels), the subhalo mass is more significantly decreased on more tightly bound (smaller $x\sub{c}$) or radial (smaller $\eta$) orbits, since the orbits have smaller pericentres and the subhalo feels a stronger tidal force from the host potential. The lower panels illustrate that when the subhaloes are on the same orbit, a larger $c\sub{h}$ (smaller $c\sub{s}$) leads to more significant mass loss because of the stronger tidal force of the host halo (more loosely bound subhalo structure). These results, which only represent a tiny fraction of the entire DASH library, are consistent with intuition.

Less intuitive is the behavior in the bound mass fraction on a highly radial orbit (i.e., small $\eta$) close to a pericentric passage. In some cases, the bound mass fraction fluctuates wildly, dropping steeply, only to increase again immediately thereafter (see for instance the purely radial, $\eta=0$ orbit in the upper-left panel). This arises i) because the method to compute the bound mass and orbit of the subhalo allows re-binding of particles (see Appendix A of \paperi{} for details), and  i\hspace{-.1em}i) because the subhalo is impulsively heated by tidal shocking, especially on radial orbits \citep[e.g.,][]{Spitzer1987, Gnedin1999}, which leads to the strong reduction of the bound mass. A fraction of the temporarily evaporated subhalo mass is re-bound during the subsequent re-virialization process of the subhalo.

Another interesting feature apparent in \autoref{fig:fb} is that, except for the step-like behavior near pericentre, $f\sub{b}$ roughly behaves like a power-law function for $t/T\sub{r}>1$. The slope of this power-law function clearly depends on the orbital and concentration parameters, and the DASH library, which contains simulation data for a large parameter space, can be used to calibrate these scaling relations, which in turn can be used to model the tidal evolution of substructure in (semi-)analytical models of structure formation. An alternative method, which we explore in \S\ref{sec:ml} below, is to use machine learning algorithms, such as random forest regression, to process and distill the huge amount of data available in the DASH library. As we demonstrate, this allows for reasonably accurate predictions for the bound mass fraction of subhaloes as a function of their time since accretion.

%%%%%%%%%%%%%%%%%%%%%%%%%%%%%%%%%%%%%%%%%%%%%%%%%%%%%%%%%%%%%%%%%%%%%%%%%%%%%%%%%%%%
\subsection{Radial profiles}
\label{ssec:profs}

\begin{figure}
\begin{center}
\includegraphics[width=0.45\textwidth]{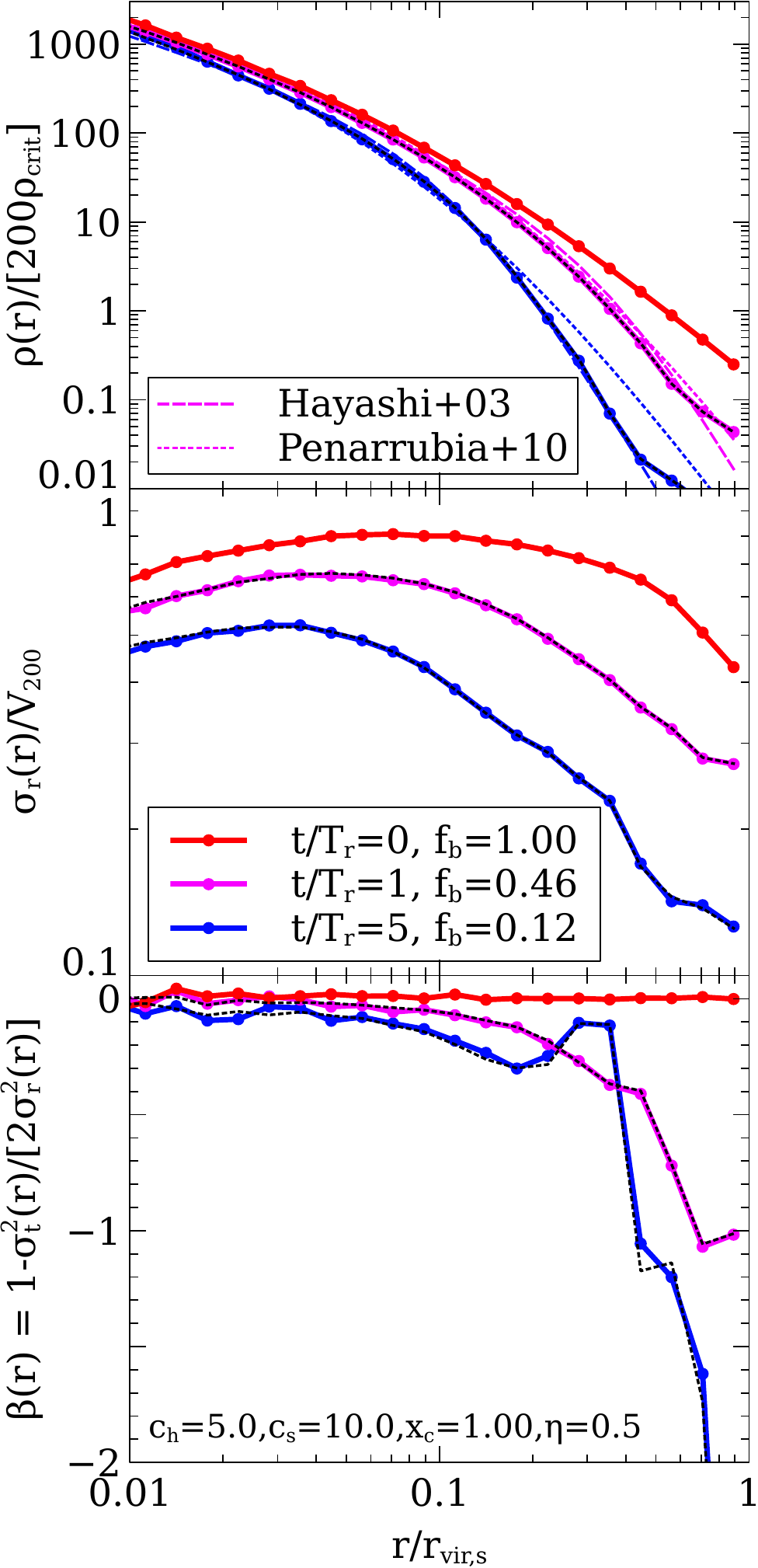}
\end{center}
\caption{
Radial profiles of the dark matter subhalo in the DASH simulation with $c\sub{h}=5.0$, $c\sub{s}=10.0$, $x\sub{c}=1.0$, and $\eta=0.5$. In each panel, coloured solid lines are the results from the simulation with the fiducial resolution and the black dashed lines demonstrate those from the higher resolution run (in which the number of particles is increased by a factor of 10, holding the other parameters fixed). ({\it Top}) Density profile. In addition to the simulation results, predictions by the models of \citet[][coloured dashed]{Hayashi2003} and \citet[][coloured dotted]{Penarrubia2010} are shown. ({\it Middle}) Profile of the radial velocity dispersion, $\sigma\sub{r}(r)$. ({\it Bottom}) Profile of the velocity anisotropy parameter, $\beta(r) \equiv 1 - \sigma^2\sub{t}(r)/[2 \sigma^2\sub{r}(r)]$ \citep[e.g.,][]{Binney2008}, where $\sigma\sub{t}(r)$ is the profile of the tangential velocity dispersion of the subhalo. The corresponding time and subhalo bound mass fraction are denoted in the middle panel.
\label{fig:radprof}}
\end{figure}

The DASH library also includes density and velocity dispersion profiles of tidally stripped subhaloes. An example is shown in \autoref{fig:radprof}, which plots the profiles for a subhalo for three snapshots of the simulation with $c\sub{h}=5.0$, $c\sub{s}=10.0$, $x\sub{c}=1.00$ and $\eta=0.5$. Each of these snapshots corresponds to an epoch at which the subhalo is near its apocentre, at which point the subhalos is in a fairly relaxed state \citep[e.g.,][]{Aguilar1986, Penarrubia2009}, and the corresponding bound mass fractions are indicated in the middle panel.  The top panel of \autoref{fig:radprof} shows the subhalo's density profile, normalized by $200 \rho_{\rm crit}$. Note how tidal stripping mainly removes mass from the outskirts, while leaving the central densities almost unaffected. For comparison, the dashed and dotted curves are the model predictions of \citet{Hayashi2003} and \citet{Penarrubia2010}, respectively. Both models suggest that the density profile of a stripped subhalo only depends on the initial density profile (prior to stripping) and the present bound mass fraction. Whereas the model by \citet{Hayashi2003} fits the profiles extremely well, the model by \citet{Penarrubia2010} predicts a shallower outer density profile at later times. However, as we demonstrate in Green et al. (in preparation), neither the \citet{Hayashi2003} nor the \citet{Penarrubia2010} model can adequately describe the evolution of the subhalo density profile under all conditions encountered in the DASH library, and we therefore develop a new and improved model based on the entire set of over 2,000 DASH simulations. The middle and bottom panels of \autoref{fig:radprof} show the radial velocity dispersion profiles and the corresponding profiles of the velocity anisotropy parameter
\begin{equation}
\beta(r) \equiv 1 - \frac{\sigma^2\sub{t}(r)}{2 \sigma^2\sub{r}(r)}, 
\end{equation}
\citep[][]{Binney2008}. Note that, by construction, the initial subhalo is isotropic ($\beta = 0$) throughout. At later times, as its bound mass fraction decreases, the subhalo becomes more and more tangentially anisotropic ($\beta < 0$) in its outskirts, while the radial velocity dispersion profile decreases on all scales. Hence, the bound remnant becomes colder and colder as more and more mass is stripped, and since particles on more radial orbits reach larger apocentric distances, they are more likely  stripped, thereby causing the remnant to become tangentially anisotropic. Finally, the black, dashed lines show the results from a simulation with 10 times higher mass resolution. The fact that the resulting profiles are indistinguishable from those of the nominal resolution simulation indicates that the DASH simulations are well converged.

%%%%%%%%%%%%%%%%%%%%%%%%%%%%%%%%%%%%%%%%%%%%%%%%%%%%%%%%%%%%%%%%%%%%%%%%%%%%%%%%%%%%
%%%%%%%%%%%%%%%%%%%%%%%%%%%%%%%%%%%%%%%%%%%%%%%%%%%%%%%%%%%%%%%%%%%%%%%%%%%%%%%%%%%%
\section{An application of machine learning to the DASH library}
\label{sec:ml}

The DASH library is a homogeneously structured, large dataset that can easily be explored with machine learning (ML). As a first example, we apply a commonly-used ML regression method, random forests \citep[RF;][]{Breiman2001} as implemented in {\sc scikit-learn} \citep{Scikit-learn}\footnote{https://scikit-learn.org}, to predict $f\sub{b}$ as a function of five features: $t/T\sub{r}$, $x\sub{c}$, $\eta$, $c\sub{h}$ and $c\sub{s}$. Because $c\sub{h}$, $c\sub{s}$ and $x\sub{c}$ are equally binned in logarithmic space in the DASH library (see \S\ref{ssec:orb_c_params}), we adopt $\log{(c\sub{h})}$, $\log{(c\sub{s})}$ and $\log{(x\sub{c})}$ as the actual features. Furthermore, we train the model to predict $\log{(f\sub{b})}$ as the target since $f\sub{b}$ varies over several orders of magnitude. In the following analysis, only data points satisfying the numerical criteria of \paperii{} are included (a brief summary is given in \S\ref{ssec:data_str}).

The ML algorithm we adopt, RF, is based on decision trees \citep{Quinlan1986}. While the decision tree method is intuitive and useful, trained models tend to be overfitted, i.e., the training data set is very accurately reproduced while poor predictions are made for untrained cases. In order to avoid this overfitting issue, RF constructs an ensemble of decision trees and adopts the mean prediction of individual decision trees as the final prediction. The ensemble consists of 20 decision trees with a maximum depth of 20. Here, the depth of the tree is the number of layers from a root to a leaf. The other hyperparameters are set to their default values in the {\sc scikit-learn} implementation. 

In order to increase the confidence in our trained model, we adopt group $k$-fold cross validation \citep[e.g.,][and references therein]{Browne2000}. First, the full data set is divided into $k$ subsets composed of data points from randomly selected simulations. Note that all time steps from each simulation are assigned to one group such that they are all placed within the same subset and thus are not split between multiple subsets. Then, the RF is trained with $k-1$ subsets and the trained model is tested on the remaining subset. This training and test procedure is iterated $k$ times and the performance ($R^2$ score, see below for details) is measured as the average of the $k$ models. We set $k = 5$ and the number of simulations in each subset is almost the same (435 or 436). To further verify the robustness of our trained model, we perform a test with a reduced number of $t/T\sub{r}$ bins. While the other four parameters ($c\sub{h}$, $c\sub{s}$, $x\sub{c}$, and $\eta$) have 11 bins for each, $t/T\sub{r}$ has 301 bins in the DASH library. This might artificially weight $t/T\sub{r}$ more strongly relative to the other features. To assuage this concern, we also train a model in which we use only 11 of the 301 epochs, equally spaced in time, such that all five features have 11 bins.

\begin{figure}
\begin{center}
\includegraphics[width=0.48\textwidth]{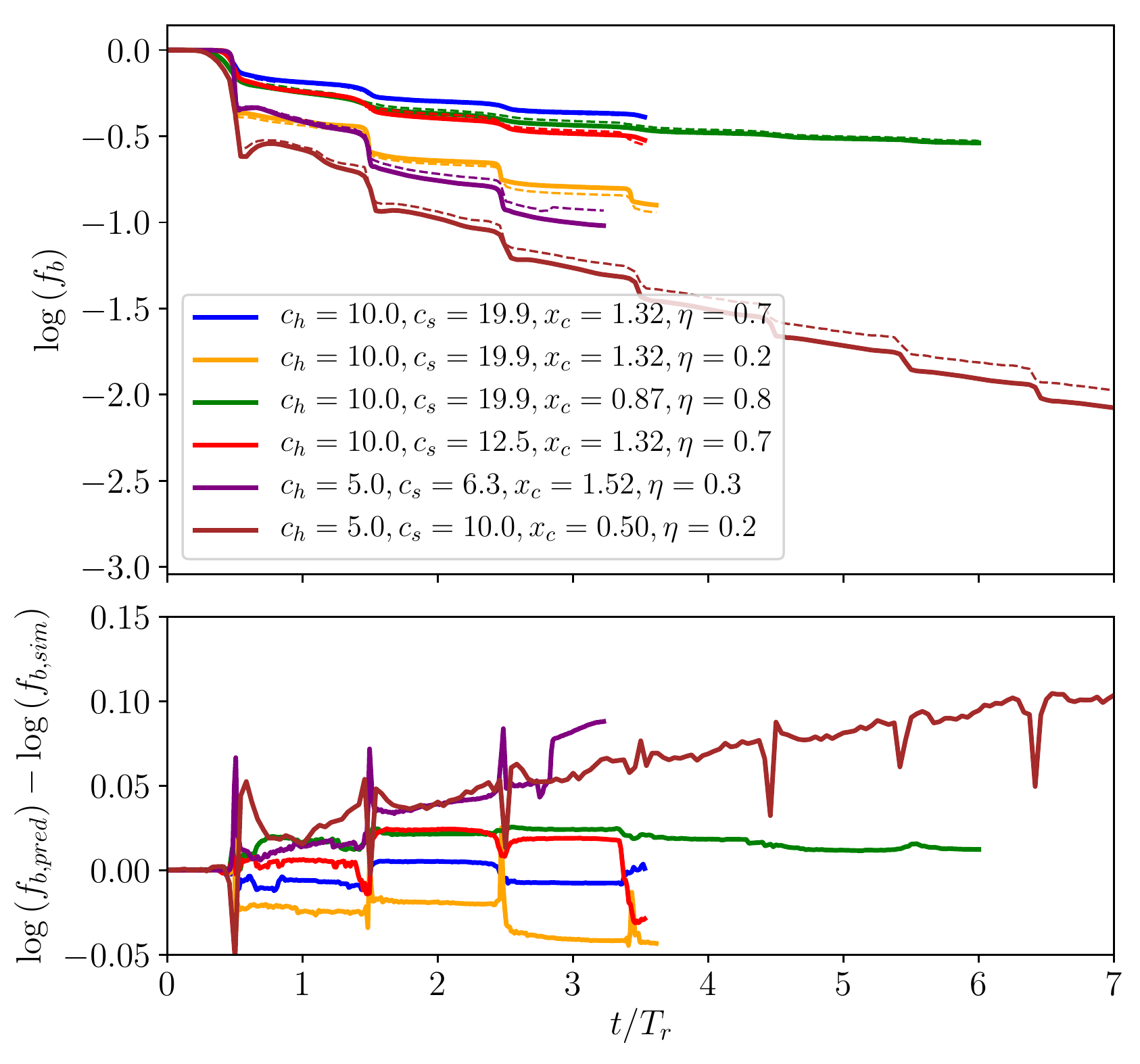}
\end{center}
\caption{
Comparison between the subhalo bound mass fraction in the simulations, $f\sub{b,sim}$, and those predicted by the model based on a commonly-used machine learning method, random forest regression, $f\sub{b,pred}$. ({\it Upper}) Solid lines represent the simulation results and dashed lines represent the prediction by the model. ({\it Lower}) Residuals between the simulations and model predictions. The same colour scheme is used as the upper panel. The times are scaled by the radial period, $T\sub{r}$.  
\label{fig:fb_ml}}
\end{figure}
\autoref{fig:fb_ml} compares the evolution of $f\sub{b}$ in the $N$-body simulations with the predictions made by the trained RF model and depicts that the model is predictive at the $\sim$\,0.1\,dex level. Note that the model predictions shown in \autoref{fig:fb_ml} are of simulations included only in the test fold for this particular trained model. In our case, a measure of accuracy of the trained model, the $R^2$ score (coefficient of determination) is defined as follows:
\begin{equation}
R^2 \equiv 1 - \frac{\sum [\log{(f\sub{b,sim})} - \log{(f\sub{b,pred})}]^2}{\sum [\log{(f\sub{b,sim})} - \langle \log{(f\sub{b,sim})}\rangle]^2}.
\end{equation}
The summation runs over all data points in the test set. The mean value in the test set is simply given as
\begin{equation}
\langle \log{(f\sub{b,sim})}\rangle = \frac{1}{N\sub{test}} \sum \log{(f\sub{b,sim})}.
\end{equation}
Here, $N\sub{test} = N\sub{snap} \times N\sub{sim}/5 \approx 131,000$ is the number of data points in the test set and $N\sub{sim}(=2,177)$ and $N\sub{snap}(=301)$ represent the number of simulations in the DASH library and the number of snapshots in each simulation, respectively. We emphasize again that only the data points fulfilling the numerical convergence criteria (see \paperii{} and \S\ref{sssec:analysis}) are included in this analysis. 
The trained model yields $R^2 > 0.98$ for the test set in all five cross-validation cases, indicating that the trained model works well for untrained cases within the covered parameter space. The $R^2$ score does not change even if we reduce the number of $t/T\sub{r}$ bins by a factor of 30, verifying that the higher number of bins in $t/T\sub{r}$ does not matter for our model. 

{\sc scikit-learn} also reports the importance of features, i.e., how much the model depends on each feature for predicting $f\sub{b}$. The derived importance for each feature is [$t/T\sub{r}$, $\log{(c\sub{h})}$, $\log{(c\sub{s})}$, $\log{(x\sub{c})}$, $\eta$] = $[0.46,0.05,0.21,0.09,0.19]$, meaning that $f\sub{b}$ depends strongly on time while $c\sub{s}$ and $\eta$ play the most dominant roles among the four given parameters. This information may be useful in constructing semi-analytical models and in determining the parameter sets to be studied in the subsequent expansion of the DASH library.

The model trained with the full data set is saved in the file named \texttt{dash\_fb\_rf.joblib} and available within the Jupyter notebook, \texttt{dash\_fb\_predict.ipynb}, in the DASH library. Inputting the four parameters we vary, $x\sub{c}$, $\eta$, $c\sub{h}$, and $c\sub{s}$, as well as the time of interest, one obtains the expected trajectories of the mass evolution of tidally stripped subhaloes. While the model is accurate and easy to use, it is not robust for extrapolation beyond the sampled region of parameter space because of the nature of decision tree-based algorithms, which can only interpolate between the data points in the training set. The prediction made by RF corresponds to the mean prediction of individual decision trees and hence it does not work in the parameter space where no data points are found. Other ML algorithms, e.g., support vector machines and others based on neural networks, are needed to construct models that can be extrapolated, but these frameworks generally require more complicated data preprocessing treatment and tuning of hyperparameters than those for RF to obtain good models. 

Subsequent studies will use the other types of data available in DASH, such as the radial profiles of mass density and velocity dispersion, in order to investigate the dynamical evolution of tidally stripped subhaloes in more detail (e.g., Green et al., in prep.). 

%%%%%%%%%%%%%%%%%%%%%%%%%%%%%%%%%%%%%%%%%%%%%%%%%%%%%%%%%%%%%%%%%%%%%%%%%%%%%%%%%%%%
%%%%%%%%%%%%%%%%%%%%%%%%%%%%%%%%%%%%%%%%%%%%%%%%%%%%%%%%%%%%%%%%%%%%%%%%%%%%%%%%%%%%
\section{Summary and discussion}
\label{sec:summary}

Cosmological $N$-body simulations are the prime tool used to study the hierarchical assembly of dark matter haloes. They reveal that virialized dark matter haloes have a universal density profile \citep[e.g.,][]{Navarro1997}, and that roughly 10 percent of their mass is bound up in distinct subhaloes \citep[e.g.,][]{Ghigna.etal.98, Gao.etal.04, Giocoli.etal.10}. According to the same simulations, a large fraction of these subhaloes completely disrupt after a few orbital periods \citep[][]{Han2016, vandenBosch2017}. It has recently been argued that the majority of this disruption is artificial \citep[][]{Penarrubia2010, PaperI, PaperII}, and thus that state-of-the-art cosmological simulations still suffer from an appreciable amount of `over-merging'. Most importantly, \cite{PaperII} argued that this problem may go unnoticed in standard numerical `convergence' tests. Hence, it is prudent that we consider alternative methods to predict the abundance and demographics of dark matter substructure, which is a potentially powerful Rosetta stone for deciphering the nature of dark matter.

One alternative to numerical simulations is a semi-analytical approach that combines halo merger trees, constructed using the framework of extended Press-Schechter theory \citep[][]{Bond.etal.91}, with a treatment of the tidal evolution of subhaloes as they orbit their host. These models are not hampered by discreteness issues or limiting force resolution responsible for artificial disruption. In addition, these models are far less CPU-intensive than actual $N$-body simulations, thus allowing for an extensive exploration of parameter space. Numerous models along this line have been constructed in the past \citep[][]{Taylor2001, Taylor.Babul.04, vandenBosch2005, Penarrubia.Benson.05, Zentner2005, Diemand2007, Kampakoglou.Benson.07, Gan2010, Pullen.etal.14, Jiang2016}. Unfortunately, since we lack a purely analytical treatment of tidal stripping and heating, these models typically contain one or more `fudge' parameters. These are tuned by requiring the model to reproduce the subhalo mass functions taken from cosmological $N$-body simulations. The obvious downside of this approach is that the models thereby inherit the over-merging problems of the simulations.

In order to overcome this dilemma, we need idealized simulations that are (i) well resolved and free from artificial disruption, and (ii) optimized to allow for calibration of semi-analytical treatments of tidal stripping and heating. This paper presents the DASH library, consisting of 2,177 idealized, high-resolution ($N=1,048,576$), collisionless $N$-body simulations of individual dark matter subhaloes orbiting in the potential of a static, analytical host halo. The simulations have sufficient mass and force resolution to overcome artificial disruption (i.e., they satisfy the numerical reliability criteria given by equations~[\ref{critA}] and~[\ref{critB}]), and sample the entire parameter space of orbital energies, orbital angular momenta, and halo concentrations relevant for dark matter substructure. All simulations adopt a host halo-to-subhalo mass ratio of $\calM = \Mhost/\Msub = 1000$, for which dynamical friction, which is not accounted for in the DASH simulations, is negligible. Because of the scale-free nature of the tidal evolution of subhaloes (see \S\ref{ssec:mass_ratio}), the DASH simulations are valid for any mass ratio large enough such that dynamical friction is negligble (i.e., $\calM \gta 100$). Each simulation is evolved for a period of roughly 36 Gyr, during which the subhalo undergoes anywhere between 2 and 12 radial orbits. For each simulation, the DASH library, which is publicly available, contains simple text files that present, among others, the temporal evolution of the subhalo's bound mass fraction, and the density and velocity dispersion profiles of the bound particles of the subhalo at 301 outputs equally spaced in time. The library also contains a Python code, trained on the DASH simulation data, that uses random forest regression to predict the bound mass fraction of subhaloes as a function of time for given halo concentrations and orbital parameters. This code, which is accurate at the 0.1 dex level, conveniently summarizes the main results from our large suite of simulations.

In a forthcoming paper (Green et al., in prep.), we use the DASH library to calibrate a new and improved semi-analytical model for the tidal evolution of subhaloes, which we will subsequently use in combination with accurate halo merger trees \citep[e.g.,][]{Parkinson.etal.08, Jiang2014} to predict the subhalo mass function of CDM haloes, unhindered by artificial disruption. This will shed new light on the level of reliability of the subhalo demographics that have been extracted from cosmological $N$-body simulations.

Finally, we emphasize that although the parameter space covered by the DASH library is vast, it is by no means exhaustive. One obvious shortcoming, as discussed above, is that the DASH simulations are inadequate to describe major mergers with $\calM \lta 100$. In those cases, dynamical friction due to the host, and self-friction due to tidally stripped material, cause the orbit of the subhalo to decay, exposing it to stronger tides.  Another degree of freedom not covered here is the inner density slope of dark matter haloes. It is well known that observations of dwarf galaxies often suggest that their haloes have constant density cores, rather than the steep $r^{-1}$-cusps predicted by dark matter-only simulations \citep[e.g.,][]{Burkert1995, Gentile2004, Oh2011, Hayashi2015}. Such cores can be created within the CDM paradigm by a variety of baryonic processes \citep[e.g.][]{El-Zant2001, Inoue2011, Pontzen2012, Ogiya2014}, and have a dramatic impact on the tidal evolution of subhaloes \citep{Penarrubia2010, Errani2015, Ogiya2018}. In addition, baryons modify the potentials of host- and subhaloes through the bulges and discs that they form at the halo centres, and these also strongly impact the tidal fields \citep{Errani2017, Garrison-Kimmel2017}. Finally, in the DASH simulations presented here, the host halo is assumed to be spherically symmetric, which allows us to completely specify each orbit with only two parameters (energy and angular momentum). Cosmological simulations, though, indicate that dark matter haloes are expected to be triaxial systems \citep[e.g.,][]{Jing2002, Allgood2006, Hayashi2007}, consistent with the shapes of the gravitational potentials of galaxies and clusters as inferred from a variety of observations \citep[e.g.,][]{Oguri2005, Corless2007, Law2010}. Triaxial systems have a much richer variety of orbits, which is likely to impact the tidal evolution of subhaloes.

In the near future, we therefore anticipate augmenting the DASH library with a suite of simulations that probe some of this extended parameter space. These additional simulations will be particularly useful for informing the semi-analytical treatments mentioned above. Ideally, any such semi-analytical treatment should capture the actual physics of tidal stripping and heating, and should thus be able to correctly predict the tidal evolution of subhaloes in triaxial potentials, in the presence of orbital decay due to dynamical friction, or in the case where the potential of the host- and/or subhalo has been modified due to the impact of baryonic processes. It remains to be seen to what extent the models can meet this challenge, and it is our hope that the DASH library presented here, as well as its future extensions, will play an important role in this process.

%%%%%%%%%%%%%%%%%%%%%%%%%%%%%%%%%%%%%%%%%%%%%%%%%%
%%%%%%%%%%%%%%%%%%%%%%%%%%%%%%%%%%%%%%%%%%%%%%%%%%
\section*{Acknowledgements}

We are grateful to the developers of {\sc scikit-learn} for making their code publicly available. GO and OH acknowledge funding from the European Research Council (ERC) under the European Union's Horizon 2020 research and innovation programme (grant agreement No. 679145, project `COSMO-SIMS'). FvdB is supported by the National Aeronautics and Space Administration through Grant No. 17-ATP17-0028 issued as part of the Astrophysics Theory Program, and by the US National Science Foundation through grant AST 1516962.



\bibliographystyle{mnras}
\bibliography{tstrip} 


%%%%%%%%%%%%%%%%%%%%%%%%%%%%%%%%%%%%%%%%%%%%%%%%%%

%%%%%%%%%%%%%%%%% APPENDICES %%%%%%%%%%%%%%%%%%%%%

\appendix

\section{Simulation Invariance}
\label{App:invariance}

As discussed in \S\ref{sssec:ICs}, the DASH simulations have been run in model units for which $G = \Msub = \rsub = 1$. In these units, the subhalo has a crossing time $t_{\rm cross} \equiv \rsub/\Vsub$ of unity. Converting to physical time units, simply requires multiplying the model time units by $1.44 \, \Gamma \Gyr$ with $\Gamma = \Gamma(h, \Delta_{\rm vir})$ given by Equation~(\ref{gammascaling}). Hence, each DASH simulation is applicable to any combination of $\Delta_{\rm vir}$ and $H_0 = 100 \, h \kmsmpc$, each of which with its own scaling between time in model units and time in physical units. 

Changing $\Delta_{\rm vir}$ and/or $h$, while keeping the simulation parameters ($\calM$, $x_\rmc$, $\eta$, $c_\rmh$, $c_\rms$) fixed, corresponds to changing the actual physical densities. For example, increasing $\Delta_{\rm vir}$ implies that the virial radius of the host halo decreases, such that a given value of $x_\rmc$ corresponds to a smaller physical radius, where the density of the host halo is larger. But, since the densities of the subhalo change similarly, and since the tidal evolution only depends on the ratio of densities (i.e., gravity is scale-free), the outcome of the simulation is entirely invariant to these changes in  $\Delta_{\rm vir}$ and/or $h$.

However, when changing  $\Delta_{\rm vir}$ and/or $h$ one can also re-scale the DASH simulations in another way, one that keeps the physical densities, and hence the mapping between model time and physical time, invariant. This scaling, however, requires a mapping between the parameter set $\{\calM, x_\rmc, \eta, c_\rmh, c_\rms\}$ for the DASH simulation ($\Delta_{\rm vir} = 200$ and $h = 0.678$) and another parameter set $\{\calM', x'_\rmc, \eta', c'_\rmh, c'_\rms\}$ corresponding to $\Delta'_{\rm vir}$ and $h'$.  This scaling keeps the characteristic density and scale-radius, i.e., $\rho_0$ and $r_\rms$ in equation~(\ref{NFWprof}), invariant. Consequently, a different value of $\Delta_{\rm vir}$, which corresponds to a different virial radius, now implies a different value for the mass and concentration parameter of the halo. And since the orbital radius remains invariant, also the dimensionless parameter $x_\rmc$ will change its value. Using that the characteristic density of an NFW halo with concentration parameter $c$ is given by
\begin{equation}
\rho_0 = \frac{c^3}{f(c)} \, \Delta_{\rm vir} \rho_{\rm crit},
\end{equation}
it is straightforward to show that a DASH simulation for $\{\calM, \chost, \csub, \xc, \eta\}$ can be used to represent the evolution of a subhalo with parameters $\{\calM', c'_\rmh,c'_\rms, x'_\rmc, \eta'\}$ for any other combination of $\Delta'_{\rm vir}$ and $\rho'_{\rm crit}$ using the following mapping:
\begin{equation}\label{Qmap}
\begin{array}{rcl}
\calM  & \rightarrow & \calM' = \calM \, (\calQ^3_\rmh/\calQ^3_\rms) \\
\chost & \rightarrow & c'_\rmh = \chost \, \calQ_\rmh \\ 
\csub  & \rightarrow & c'_\rms = \csub  \, \calQ_\rms \\
x_\rmc & \rightarrow & x'_\rmc = x_\rmc \, \calQ^{-1}_\rmh \\ 
\eta   & \rightarrow & \eta' = \eta \\
f_\rmb & \rightarrow & f'_\rmb = f_\rmb \, (\Gamma/\calQ^3_\rms)\,.
\end{array}
\end{equation}
Here $\Gamma = \Gamma(h',\Delta'_{\rm vir})$ is given by Equation~(\ref{gammascaling}), and $\calQ = \calQ(c,\Delta'_{\rm vir})$ is the root of
\begin{equation}
\calQ^3 \, \frac{f(c)}{f(c \, \calQ)} \, \left(\frac{\Delta'_{\rm vir}}{200}\right) \, \left(\frac{h'}{0.678}\right)^2 = 1\,.
\end{equation}

Note that, whereas the $(\Delta_{\rm vir},h)$-dependent time-scaling is exact, this density-invariant mapping is only approximate. One of the reasons is that, because of dynamical friction, our simulations are only invariant to changes in $\calM$ as long as the mass ratio $\Mhost/\Msub \gta 100$. Hence, one can only use this mapping as long as $(\calQ_\rmh/\calQ_\rms)^3 \gta 0.1$. In addition, the initial subhalo is only initialized out to a truncation radius $r_{\rm trunc}$ (see \S\ref{sssec:ICs}), which is equal to the virial radius, but only for $\Delta_{\rm vir} = 200$ and $h = 0.678$. For other values of $\Delta_{\rm vir}$ and $h$, the subhalo is effectively truncated at a radius $r_{\rm trunc} =  r'_{\rm vir}/\calQ_\rms$. As we have demonstrated in \paperii{}, the  simulation outcome depends only very weakly on where exactly the initial subhalo is truncated, as long as it is outside of the initial tidal radius. The weak dependence mainly originates from `self-friction' (see \S\ref{ssec:mass_ratio}), which depends on the amount of mass that is stripped from the subhalo, which is larger if the initial truncation radius is larger. Hence, as long as the impact of self-friction is weak, which is almost always the case, we expect the density-invariant mapping of Equation~(\ref{Qmap}) to be reasonably accurate. We have verified that this is indeed the case by running a few simulations for different values of $\Delta_{\rm vir}$, and comparing the resulting $f_\rmb(t)$ to predictions from the re-scaled DASH simulations based on the mapping of Equation~(\ref{Qmap}). Except for the time period prior to the first pericentric passage, we find this mapping to be accurate at the few percent level. The reason why the mapping fails prior to the first-pericentric passage is simply that the mapping predicts that the initial subhalo starts out with a bound mass fraction $f'_\rmb$ that is not equal to unity. However, after first pericentric passage the subhalo is basically stripped down to the same physical radius as in the fiducial case, and the mapping of Equation~(\ref{Qmap}) is reliable.

%%%%%%%%%%%%%%%%%%%%%%%%%%%%%%%%%%%%%%%%%%%%%%%%%%

% Don't change these lines
\bsp	% typesetting comment
\label{lastpage}
\end{document}